\documentclass[10pt, reprint, aps, amsmath,amssymb,twocolumn]{revtex4-1}

\usepackage{epigraph}
\usepackage{graphicx}
\usepackage{dcolumn}
\usepackage{bm}
\usepackage[font=scriptsize]{caption}
\usepackage{booktabs}
\usepackage{subcaption,graphicx}

\begin{document}

\title{Characterizing Complementary Bipolar Junction Transistors by Early Modelling, Image Analysis, and Pattern Recognition}

\author{Luciano da F. Costa}
\email{ldfcosta@gmail.com}
\affiliation{S\~ao Carlos Institute of Physics, IFSC-USP,  S\~ao~Carlos, SP,~Brazil}

\date{\today}

\begin{abstract}
This work reports an approach to study complementary pairs of bipolar junction transistors, often used in push-pull circuits typically found at the output stages of operational amplifiers. After the data is acquired and pre-processed, an Early modeling approach is applied to estimate the two respective parameters (the Early voltage $V_a$ and the proportionality parameter $s$). A voting procedure, inspired on the Hough transform image analysis method, is adopted to improve the identification of $V_a$. Analytical relationships are derived between the traditional parameters current gain ($\beta$) and output resistance ($R_o$) and the two Early parameters. It is shown that $\beta$ tends to increase with $s$ for fixed $V_a$, while $R_o$ depends only on $V_a$, 
varying linearly with this parameter. Several interesting results are obtained with respect to 7 pairs of complementary BJTs, each represented by 10 samples. First, we have that the considered BJTs occupy a restricted band along the Early parameter space, and also that the NPN and PNP groups are mostly segregated into two respective regions in this space. In addition, PNP devices tended to present an intrinsically larger parameter variation. The NPN group tended to have higher $V_a$  magnitude and smaller $s$ than PNP devices. NPN transistors yielded  comparable $\beta$ and larger $R_o$ than PNP transistors. A pattern recognition method was employed to obtain a linear separatrix between the NPN and PNP groups in the Early space and the respective average parameters were used to estimate respective prototype devices. Two complementary pairs of the real-world BJTs with large and small parameters differences were used in three configurations of push pull circuits, and the respective total harmonic distortions were measured and discussed, indicating a definite influence of parameter matching on the results.
\end{abstract}

\maketitle

\setlength{\epigraphwidth}{.49\textwidth}
\epigraph{\emph{``Unity and complementarity constitute reality.''}}{W. Heisenberg}

\section{Introduction}

Transistors, in their discrete or integrated versions, remain at the heart of modern electronics.  While these devices typically
operate as switches in digital electronics, they are often used as amplifiers in \emph{linear} or \emph{analog electronics}.
As implied by its own name, linear electronics usually requires these amplifiers to be as much linear as possible, in order
to avoid distortions and preserve the original signal information, except for the changes in amplitude 
magnification that constitutes the main purpose of amplification (e.g.~\cite{jaeger:1997,cordell:2011,self:2013,sedra:1998,analogdesign:1991}).  As it could be expected, the success in
achieving good linearity of transistors depends, to a large extent, on \emph{characterizing} their electronic properties, as well as
\emph{modeling} these devices, which can then be combined into circuits with improved linearity 
and other functionalities (e.g. filtering, mixing, etc.).   

The beginnings of modern electronics can be traced back to the transistor invention~\cite{riordan:1997}.  Bipolar junction transistors 
-- BJTs -- followed suit, playing a decisive role in the popularization of electronic equipments.
BJTs, still largely used nowadays, are current-controlled devices typically characterized by their
current gain $\beta$ and output resistance $R_o$.  One of the problems that initially challenged the applications of BJTs is the fact
that $\beta$ tends to vary largely among samples even from the very same lot.  Other important properties of a BJT, such as 
its output resistance $R_o$, are also subjected to variability.  Ultimately, it was the systematic application of
negative feedback (e.g.~\cite{black:1934}) that paved the way to the widespread application of BJTs in electronics.  The principle
here is to exchange gain for parameter invariance, linearity and extended bandwidth 
(e.g.~\cite{black:1934,ochoa:2016,chen:2017}), though it should be reminded that these effects are respective to each of the 
four types of electronic feedback (e.g.~\cite{black:1934,ochoa:2016,chen:2017}).  However, despite its
impressive efficacy, there are limits to what can be accomplished with negative feedback.  For instance, 
strong nonlinearities will require more intense negative feedback levels and respectively implied large gain reductions.    

One of the simplest amplifying approaches using BJTs is the common emitter circuit operating in classes A 
(e.g.~\cite{cordell:2011,boylestad:2008}), in which the operation point is often located near the center of the linear region excursion (assuming 
symmetric signals) in order to optimize linearity and signal excursion.  Unfortunately,  this circuit requires substantial constant  
DC biasing, implying in heating and energy loss.  An alternative approach, using a \emph{complementary pair} of
transistors -- one PNP and one NPN, is the \emph{push-pull} configuration (e.g.~\cite{self:2013,sedra:1998,analogdesign:1991,gray:1990, jaeger:1997, cordell:2011}), operating in B or AB classes. 
Basically, in the former case, the positive 
part of the input  signal is amplified by the NPN transistor, while the negative feeds the PNP BJT, therefore allowing null 
biasing and virtually no loss of DC power, also reducing thermal dissipation.  The push-pull configuration is particularly
important in electronics as it is adopted as output stage for many off-the-shelf operational amplifiers 
(\emph{OpAmps})~\cite{raikos:2009, tobey:1971}.

Yet, all these circuit types are not completely linear.   First, we have the fact that $\beta$ and $R_o$ vary with 
the input signal magnitude, implying nonlinearities.  Shortcomings also arise in the simplest push-pull configuration:
(i) it is prone to \emph{thermal runaway}, so that special care is required to limit the output current, (ii)
it incorporates an inherent nonlinearity occurring as the signal of input signal \emph{crosses-over} the null operation 
point, and (iii) two power supplies are now required (one positive and another negative).  Though negative feedback 
(e.g.~\cite{black:1934,ochoa:2016,chen:2017}) is typically employed in order to minimize this unwanted distortion,
a recent study~\cite{costafeed:2017}, considering a set of individual NPN BJTs operating in Class A, has suggested 
that moderate levels of negative 
feedback may not be enough to guarantee total BJT parameters invariance.   Therefore, proper negative feedback 
action may benefit from starting with a good linearity level.

Though a better understanding of the operation of BJT as building blocks can contribute to devising better circuit configurations,
BJTs turn out to be relatively complicated to characterize as a consequence of quantum mechanics effects, variability of fabrication
parameters, contamination, among other issues.   One of the most interesting manners to better understand a BJT
(as well as any other transistor) is by using some \emph{modeling} approach.  A particularly simple,
geometrical model based on the Early effect~\cite{costaearly:2017} was recently proposed that allows an intuitive 
representation of
the transistor transfer curve (or function) by using only two parameters:  the Early voltage ($V_a$) as well as the
proportionality parameter $s$ (it has been experimentally verified in~\cite{costaearly:2017} and in this work that,
usually, $\theta \approx s I_B$).   While traditional parameters
such as $\beta$ and output resistance $R_o$ vary with the collector voltage ($V_c$) and current  ($I_c$), the two 
Early model parameters, $V_a$ and $\beta_\theta$ remain mostly constant (inside the linear operation region) for a 
given BJT.  Thus, in a sense the Early parameterization seems to be more intrinsic and natural to BJTs.

The main purpose of the present work is to address the questions of modeling and characterization of the properties of
complementary small signal transistors.  Starting with experimental data corresponding to the voltages and currents 
of complementary
pairs of BJTs subjected to DC scanning, we then applied the Early approach so as to map each 
device into the \emph{Early space}, defined by the respective Early voltage ($V_a$) 
and the proportionality parameter $s$.  However, unlike in~\cite{costaearly:2017}, $V_a$ estimation is herein performed 
by using a voting scheme inspired on the image analysis technique known as the Hough transform
(e.g.~\cite{hough:1962,costa:1993,shapebook}).  By reducing the interference between the isoline projections into the $V_c$ 
axis, while emphasizing the coherent isolines, this methodology
has  potential for improving the accuracy for $V_a$ and $s$ estimation.  Analytical relationships are also established
between the Early parameters $V_a$ and $s$ and the more traditional current gain $\beta$ and output resistance $R_o$ 
parameters of BJTs.  It is found that the average values of $\beta$, namely $\langle \beta \rangle$  tend to increase
with $s$ for fixes $V_a$.  It is also verified that the output resistance $R_o$ increases linearly with the magnitude
of $V_a$ and is completely independent of $s$.   In principle, however, the relationship between $V_a$ and $s$ was 
unknown, but the 
reported experimental results demonstrated that they tend to vary so as to keep $\beta$ within a relatively narrow range.
These results also corroborated the linear relationship between $\theta$ and $s$, namely $\theta = s I_B$ ($I_B$
is the base current) for all considered devices, as had been suggested in a previous work~\cite{costaearly:2017}.  
The estimation errors for $V_a$ and $s$
were found to be relatively small, being somewhat larger for the PNP cases than the NPN as a consequence of the
larger parameter variation of the former group.  The distributions of $V_a$, $s$, $\langle \beta \rangle$, $R_o$ and the
amount of votes obtained for each isoline in the $V_c X I_C$ space were estimated and showed several interesting
features.  Two other parameters,
corresponding to the regression offset while estimating the isolines as well as the relationship between
$\beta$ and $I_b$ were also discussed in terms of the respective densities.  The pairwise relationship between
$V_a$ and $s$ resulted in two well-separated groups corresponding to the NPN and PNP transistors in both the
Early and the $\langle \beta \rangle \times \langle R_o \rangle$ spaces.  Several
additional interesting results were obtained, including the larger parameter variation of PNP devices and other
differentiating properties.   A  pattern recognition method, namely linear discriminant analysis (LDA), was then
applied in order to identify an optimal linear separation boundary between the two types of transistores, and the
respective average parameters were used to estimate prototypical configurations for the NPN and PNP devices.
An analytical development considering the derived equations with respect to total harmonic distortion in 
simplified common emitter configurations showed that the THD for the considered circumstances does not
vary with $V_a$, being a function of $s$ only.
In order to illustrate the application potential of the reported methods and results, three pairs with varying
parameter similarity were used to build three push pull circuit configurations, and the results confirm the importance
of parameter matching for reducing the THD level.  

This article starts by briefly presenting the three considered push pull configurations using complementary pairs of
BJT, and proceeds by describing the data acquisition system, the experimental procedure, and pre-processing. 
The Early effect, voltage and modeling is then presented, together with equations for relating the Early
parameters locally with the more traditional parameters $\beta$ and $R_o$.  However, these expressions
can only characterize these relationships locally, for specific $(V_c,I_c)$ settings.   They are here
extended to express relationships between $V_a$ and $s$ and averages of $\beta$ and $R_o$ estimated
through the whole operating region in the  $(V_c,I_c)$ space, allowing a direct bridge between these two
alternative parametric representations.
The enhanced method for Early parameter estimation, based on Hough transform voting, is then described,
illustrated, and discussed. 
The obtained results are reported and discussed next, first by individual densities, then in pairwise fashion
between parameters.  The definition of a prototypical Early space and NPN/PNP groups is also presented.
Two complementary pairs with varying levels of parameter similarity are then employed in three push
pull circuit configurations and the respective THDs are measured and discussed.  This article concludes
by recapitulating its main contributions and relating several prospects for future investigations.

\section{Three Push-Pull Configurations} \label{sec:pushpull}

\begin{figure}[h!]
\centering{
\includegraphics[width=8cm]{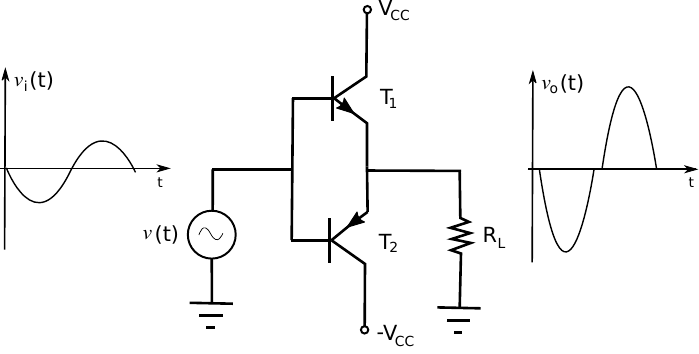}
\caption{One of the simplest push-pull configurations.  The input signal $v_i(t)$ is driven into the bases of the two transistors $T_1$ 
and $T_2$.  The positive portion of this signal is amplified by $T_1$, while the negative is processed by $T_2$.  
The outputs of these two complementary transistors are added up, feeding the load resistance $R_L$.  Observe the 
two required symmetric power supplies and that cross-over distortion appears around the null voltage of the output signal. 
Note that this
is not a practical circuit as, among other issues, it does not incorporate current limitation at the output and is prone to thermal runaway. Also,
it would be needed to decouple DC current to the input and output lines through respective capacitors.}
\label{fig:pushpull}}
\end{figure}

Figure~\ref{fig:pushpull} depicts the simplest push-pull circuit including a complementary pair of transistors (PNP and NPN),
marked as $T_1$ and $T_2$.  The input signal $v_i(t)$ enters the bases of these two transistors, which have their collector-base 
junctions inversely biased and the base-emitter junctions directly biased.  The positive portion of the input signal is
amplified at $T_1$, and the negative part is dealt with by $T_2$.  The outputs of these two transistors are linearly
superimposed onto the resistive load ($R_L$), adding up the positive and negative parts of the amplification.  In a sense,
the two complementary transistors act in integrated fashion, almost as a single device.  Two power supplies are required, 
typically $V_{CC}$  and $-V_{CC}$.   Observe that cross-over noise is implied by the offset voltages (about $0.6V$) induced 
by the forward biased base-emitter junctions.   It is interesting to notice that the push-pull configuration can be thought of as a combination 
of two complementary common emitter circuits biased at null voltage, but with the difference that the output
sign is inverted as a consequence of deriving it from the emitters rather than the collectors.  Observe that the currents entering
the collectors have not other option (except for effects such as current leakage) than leaving the transistor and flowing through the
load resistance. As in single transistor 
common emitter configurations, negative feedback (e.g.~\cite{black:1934,ochoa:2016,chen:2017}) will tend to increase the 
output resistance, which is  not desirable when driving reactive loads  because it can lead to phase distortion.

\begin{figure}[h!]
\centering{
\includegraphics[width=8.5cm]{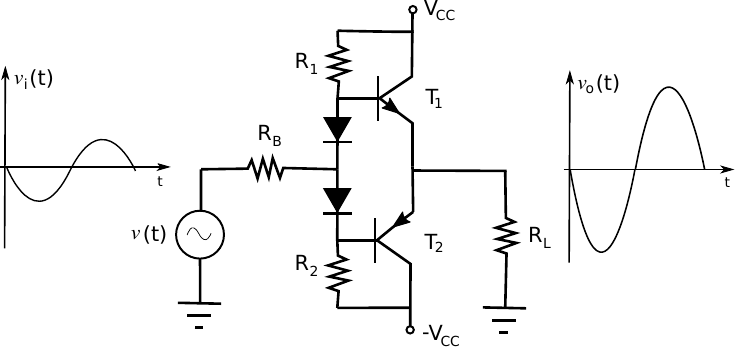}
\caption{The simple push pull configuration in Figure~\ref{fig:pushpull} can be improved by adding two diodes so as to 
compensate for the offset of the forward biased base-collector junctions of the two BJTs.  Respective resistors $R_1$ and $R_2$
are added in order to bias the diodes. Note that additional efforts are required to make this circuit practical.}
\label{fig:negpushpull}}
\end{figure}

Several means can be applied to minimize the unwanted cross-over distortion.  Two possible approaches are shown in
Figures~\ref{fig:negpushpull}, and Fig.~\ref{fig:negpushpull}.  In the first case, two diodes are added so as to compensate
the offset (about $0.6V$) implied by the forward bias of the base-collector junctions.  The resistors are added in order to bias the 
diodes.  The circuit in Figure~\ref{fig:negpushpull} includes an operational amplifier (op amp) for implementing negative
feedback.  Now, the input signal is driven into the positive lead of the op and, while the negative monitors the 
output signal at the load resistance $R_L$.  

\begin{figure}[h!]
\centering{
\includegraphics[width=6.7cm]{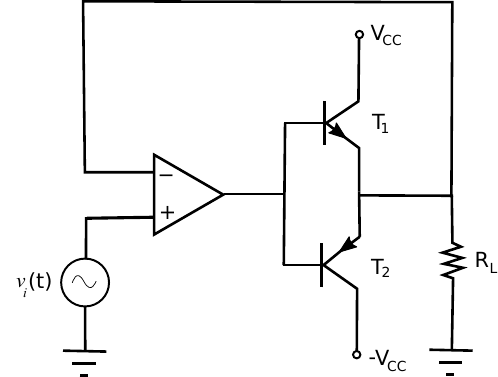}
\caption{Negative feedback version of the push pull configuration in Figure~\ref{fig:pushpull}.  The output signal
$V_o(t)$ is negatively fed back into the op amp input, contributing to make the output more similar to the input signal.  
Note that additional efforts are required to make this circuit practical.}
\label{fig:nfpushpull}}
\end{figure}

\section{The Data Acquisition System}

\begin{figure}[h!]
\centering{
\includegraphics[width=5cm]{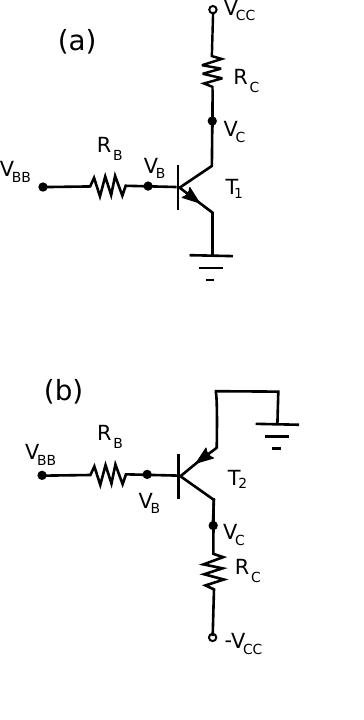}
\caption{The experimental set-up employed for imposing the voltages $V_{CC}$ and $V_{BB}$ and obtaining the
respective values of these two voltages, as well as $V_C$ and $V_B$.  Measuring back the fed voltages $V_{CC|}$
and $V_{BB}$ is adopted in order to have direct access to the voltages in the circuit, avoiding open loop
estimation of these two values. For simplicity's sake \emph{all voltages and currents in the PNP case, which
are all negative, are considered in the present article in magnitude (positive values).}}
\label{fig:acq_conf}}
\end{figure}

Each of the considered transistors was scanned by a custom-designed microprocessed system, yielding respective voltages and
currents.   The basic experimental set-up is shown in Figure~\ref{fig:acq_conf}, with respect to measurements of
NPN and PNP devices, respectively.   Recall that, in these two circuit configurations, all voltages and currents in the 
PNP are opposite to those in the NPN case.   For simplicity's sake \emph{all voltages and currents in the PNP case, which
are all negative, are used in the present article in magnitude (positive values).}.

DAC-generated voltages are applied 
at the points $V_{BB}$ and $V_{CC}$, and the voltages at these points, as well as at the points $V_B$ and $V_C$, are 
acquired by respective ADCs.  The base and collector currents can be immediately determined from these voltage values, 
taking into account Ohm's law and the values of $R_C$ and $R_B$.  The data is obtained by scanning $V_{CC}$ from $0V$ to 
$10V$ at $\Delta V_{CC}$ steps and, for each of the values, $V_{BB}$ is scanned from $0V$ to $2V$ at $\Delta V_{BB}$ steps.
The herein reported experiments considered 256 points of resolution for $V_{BB}$ and 128 points for $V_{CC}$.  
Thus, the characteristic surface $I_C \times V_C$ of  each transistor can be obtained by sliding the load line with a
load resistance $R_L = 673 \Omega$, as a consequence of the variable $V_{CC}$.  Other scanning procedures can
be used, leading to similar results as the characteristic surface of a transistor does not depend on external compontens.  
Observe that the value of $R_L$
is not critical and nearly the same characteristic surfaces, which depends only on the device and not on the attached
resistances, will be obtained for the same BJT by using different values of $R_L$, the absolute maximum ratings being
observed.

\begin{figure}[h!]
\centering{
\includegraphics[width=7cm]{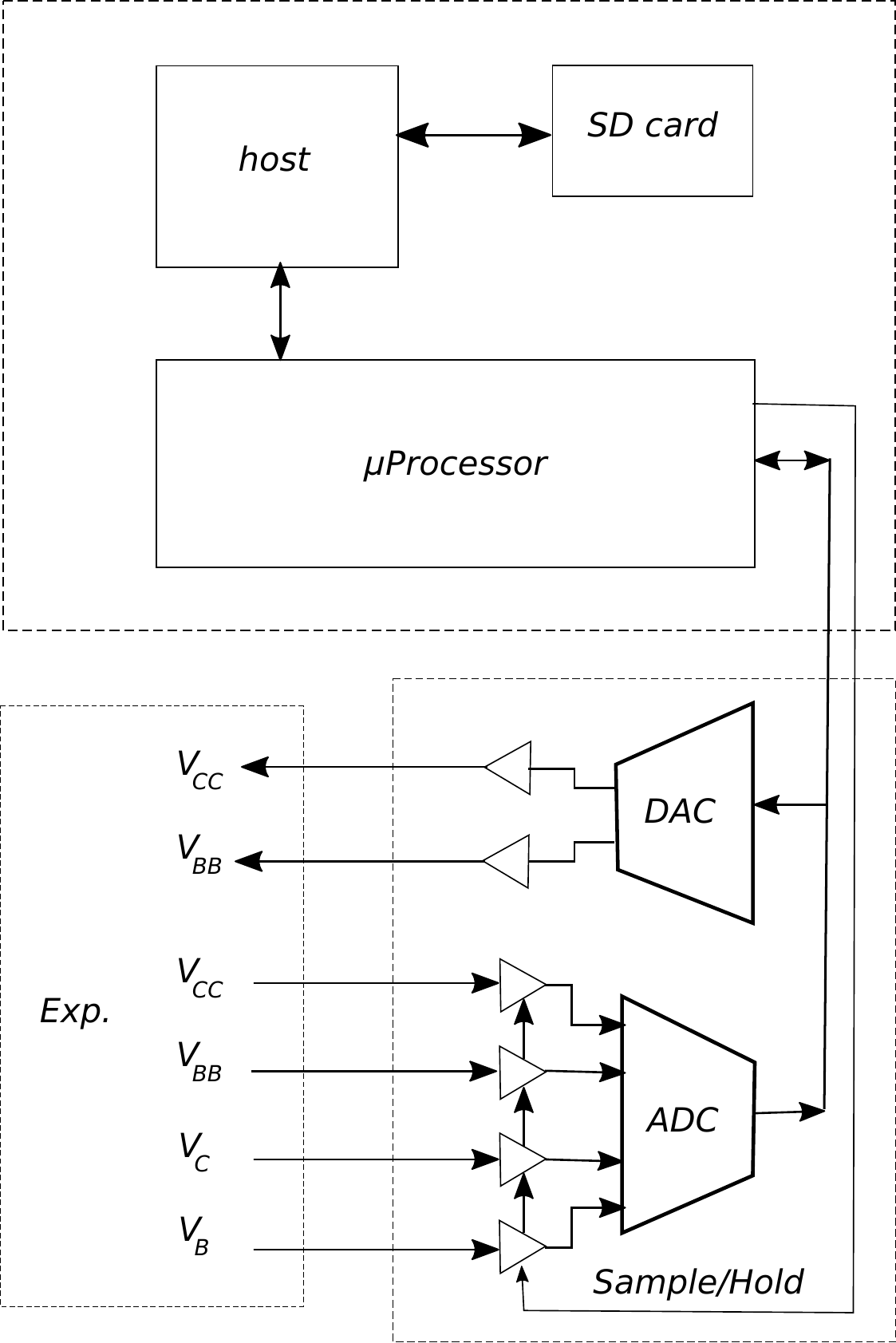}
\caption{The microprocessor-based system used for data acquisition.  The microprocessor subsystem,
which also includes memory and driving circuitry, is connected to a twin DAC and a quadruple ADC.
The DAC is buffered drivers with adjustable gain and offset.  The ADC is connected to sample-and-hole,
buffering circuits, which have their sampling signal in common so as to ensure synchronous acquisition
of the voltages $V_{CC}$, $V_{BB}$, $V_C$ and $V_B$.  The microprocessor subsystem connects to
a host for mass memory (SD) and internet connection.}
\label{fig:microsys}}
\end{figure}

The acquisition system uses a 16-bit microprocessor connected to a host providing mass memory resources (SD card)
and internet connection  The microprocessor subsystem also includes RAM and ROM memory as well as decoding and
buffering circuitry.  The microprocessed system is illustrated in Figure~\ref{fig:microsys}.   A 12-bit twin digital-analog converter 
(DAC) is used, with respective driving buffers allowing programmable gain and offset.   The signal acquisition is performed 
by a 12-bit quadruple multiplexed input analog-digital converter (ADC), and each channel is provided with separate buffering 
and sample/hold circuitry.  Observe that all sample-and-hold circuits are strobed simultaneously to guarantee a 
snapshot of the  DC voltage state of the monitored circuit, improving the coherence of the measured values. The digital and 
analog portions of this system occupy separate printed circuit boards for improved noise immunity, with the analog board 
being cased in a special shielding box.   All power supply inputs and voltage rails are thoroughly decoupled.  The observed
noise level at the experiment portion of the analog board is in the order of the smallest division of the ADCs.  The 
microprocessed subsystem connects to a host computer in order to allow internet access and mass memory (mainly
SD card), and the data is analyzed in another, separated off-the-shelf microcomputer. 

\section{Acquisition and Pre-Processing} \label{sec:acq}

The four acquired voltages, $V_{CC}$, $V_{BB}$, $V_C$, and $V_B$, as well as the derived currents
$I_C$ and $I_B$, are then used in order to obtain the isolines of the $I_C \times V_C$ transistor characteristic 
surface, which are illustrated in Figure~\ref{fig:IaXcs} .  This is done as follows:   First, the data
indices are properly reorganized in order to derived the isolines obtained respectively to each $V_{BB}$ represented
as functions of $I_B$.  Next, each isoline is slightly smoothed with an average filter along $I_B$, in order to reduce
eventual acquisition and resolution noise.  Then, a linear resampling (actually an undersampling) is applied along 
these isolines (the interpolation free variable corresponds to $I_B$) so as to obtain a fixed, angularly equispaced, 
number of isolines $N_i$ for every transistor and reinforce the quality of the signals.
As a consequence the $\Delta I_B$ step isolines from one isoline to the next is not the same for all transistors,
but the $V_C \times I_C$ is covered in a more uniform manner. We have adopted $N_i = 100$ for all reported experiments.

\section{The Early Modeling Approach}

\begin{figure}[h!]
\centering{
\includegraphics[width=8cm]{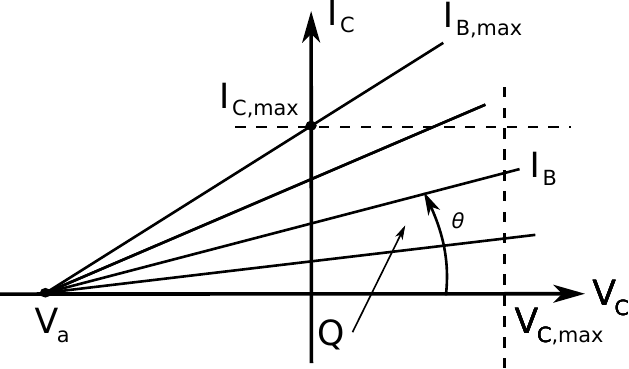}
\caption{The Early effect implies that prolongation of the isolines in the $V_C \times I_C$ space, parametrized
by $I_B$, intersect the $V_C$ axis at a well defined value $V_a$, known as \emph{Early voltage}.  The 
typical operation of a BJT, with the collector-base junction inversely biased and base-emitter junction forward 
biased, occurs within the operation region Q, observed the power, current, frequency, and voltage limitations of 
the device.}
\label{fig:Early}}
\end{figure}

Discovered by James M. Early in 1952 (\cite{early:1952}), the \emph{Early effect} corresponds to an important
phenomenon taking place inside junction transistors.  Basically, the width of space charge at the base
decreases as collector voltage is increased, yielding an increase of the current gain $\alpha$.  This
phenomenon is dominant in defining the collector conductance and is used in several BJT modes
jointly with several other parameters.   Graphically, the Early effect can be simple and neatly identified by a 
convergence of the $I_B$ isolines at a negative value $V_a$ (the \emph{Early voltage}) along $V_C$ axis, as
illustrated in Figure~\ref{fig:Early}, which also shows the involved variables and the rectangular region
of interest $Q$ (delimited by the axes and the dashed lines) in which our analysis is based.  This region
is characterized by forward biasing of the base-emitter junction and inverse bias of the collector-base
junction.  It is argued here that, to a good extent, the electronic properties of a junction transistor (e.g.~$\beta$ 
and $R_o$) in the linear region(disregarding the relatively small strips for the cut-off and saturation regions)
are to a large extent determined by $V_a$.   For instance, the larger the magnitude value of this voltage, the larger 
the output resistance $R_o$.   At the same time, larger the magnitude of $V_a$ may also affect $\beta$, 
but this depends on how the angle $\theta$ varies with $I_B$.  Henceforth, $\theta$ is given in radians values.
A relationship between $V_a$ and the values of $\beta$ average through the region $Q$ is derived further
in this article.  Observe that the Early parameterization can be understood as being more intrinsic, natural
and effective than the more traditionally used approaches, because the Early parameters $V_a$ and $s$
mostly do not vary with $V_C$ or $I_C$ in the linear region.  Yet, it is possible to derive other more traditional
parameters from the Early formulation, as done in this work.

A simple model of junction transistors by using only the Early effect has been reported recently,~\cite{costaearly:2017} which is
henceforth adopted.  This model, which is restricted to the linear region of operation, involves only two 
parameters: (i)  $V_a$ and (ii) $s$, the latter corresponding to the constant of proportionality between $I_B$ and $_\theta$
(i.e.~$\theta = s I_B$).  This linear relationship, which is henceforth adopted in this work, was observed for
several real-world BJTs~\cite{costaearly:2017} and confirmed for the BJT devices adopte in the present work.   
 Given only these two parameters, and the $I_B$ defining a isoline, 
$I_C$, indexed by $I_B$, can be easy and conveniently obtained as:

\begin{eqnarray}
    I_C = tan(\theta) (V_C - V_a) = \\ \nonumber
    = tan(s~I_B) (V_C - V_a).
\end{eqnarray}  

Observe that $tan(\theta)$ acts as a conductance, having $\mho$ (Mho, or Siemens) as unit.

Alternatively, we have that:

\begin{eqnarray}
    V_C = \frac{I_C}{tan(\theta)} + V_a \\ \nonumber
    = \frac{I_C}{tan(s~I_B)} + V_a.
\end{eqnarray}  

The value of $I_b$ for which the isolines cross the $I_C$ axis at its maximum value $I_{C,max}$ can be 
immediately calculated as:

\begin{equation}
  I_{B,max} = atan(I_{C,max} / V_a)/s,
\end{equation}  

which implies $\theta_{max} = s I_{B,max}$.

The main traditional parameters of the junction transistor --- namely the current gain $\beta$, the output resistance
$R_o$, as well as the transresistance $R_T$ --- can now be determined as:

\begin{eqnarray}
	\beta (V_C,I_C) = \frac{\partial I_C}{\partial I_B} ~\bigg\rvert_{V_C}  =  s~(V_C-V_a)~sec^2(s I_B),    \\
        R_o (V_C,I_C) =  \frac{\partial V_C}{\partial I_C} ~\bigg\rvert_{I_B}  =  \frac{1}{tan(s~I_B)}, \\
	R_T (V_C,I_C) = \frac{\partial V_C}{ \partial I_B} ~\bigg\rvert_{I_C}  =  - \frac{s~I_C}{sin^2(s~I_B)}.
\end{eqnarray}  

The simplicity and graphical intuition of these equations are a welcomed consequence of the straightforward 
Early modeling approach.

Table~\ref{fig:earlycases} shows the theoretically obtained isolines for several combinations of values of the parameters
$V_a$ and $s$.  Observe that these parameters are given for the whole first coordinate system quadrant,
ignoring the saturation and cut-off region for simplicity's sake.   In real-world transistors, these two
regions would limit, to a relatively small extent, the region of linear operation of the junction transistor respectively.  
These case-examples illustrate the behavior of the Early model parameters $V_a$ and $s$ with respect to isolines 
equispaced by $\Delta I_b = 2 \mu A$ in the $V_C \times I_c$ space for 9 combinations of values of the
$V_a = -100, -80, -20$ and $s = 3, 4, 5$ parameters.   If $s$ is kept constant (as well as the bounding region
of the $I_C \times V_C$ space), BJTs with larger $V_a$ magnitude will have larger $R_o$.   
However, it does not necessarily follow that junction transistors with larger
$V_a$ will have larger or smaller average $\beta$.  The inference of this dependency is part of the objectives of the
current work.  Actually, because the average current gain in small signal industrialized
transistors tend to be roughly similar, it could be expected that devices having larger $V_a$ magnitude will tend to have
smaller $s$ (or vice-versa), in order to keep the current gain nearly constant among devices.  The relationship between these
parameters will be further explored in the experimental sections of this article. 

\begin{table*}
  \centering
     \begin{tabular}{ccccc}
        \toprule
         \includegraphics[width=6cm]{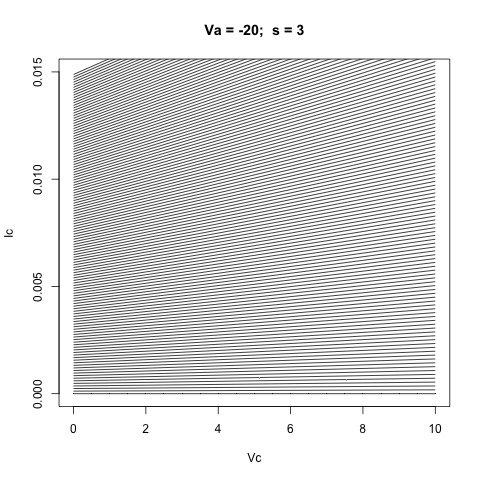} &   \includegraphics[width=6cm]{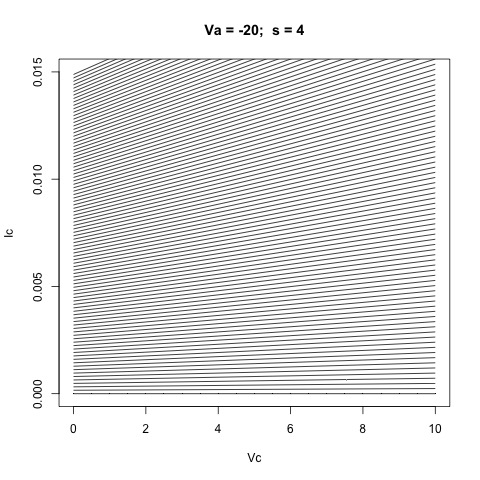} &   \includegraphics[width=6cm]{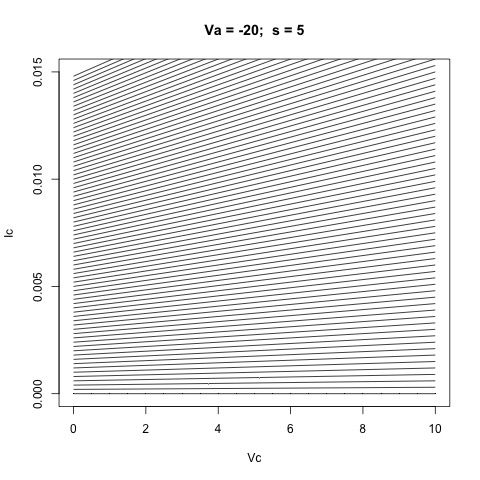}\\
         \includegraphics[width=6cm]{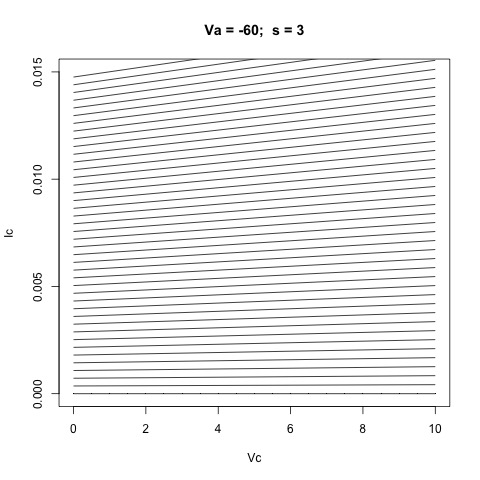} &   \includegraphics[width=6cm]{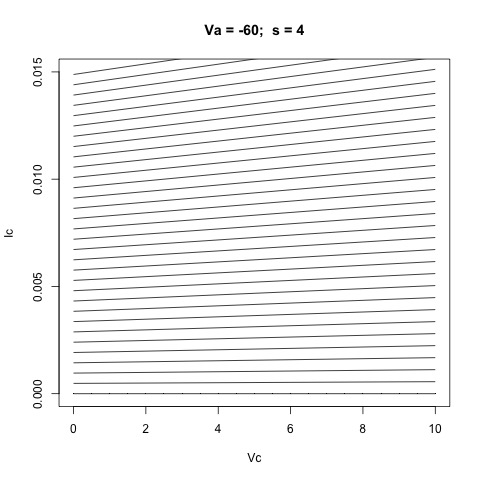} &   \includegraphics[width=6cm]{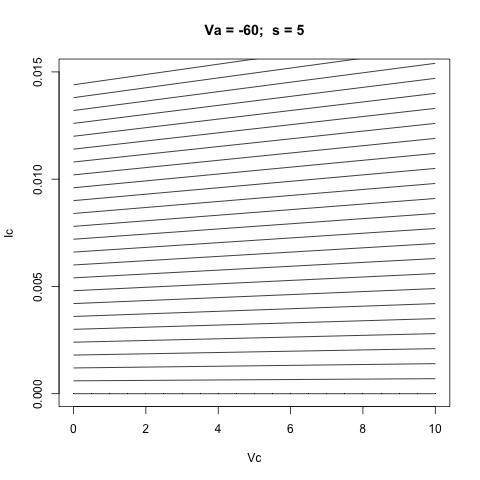}\\
         \includegraphics[width=6cm]{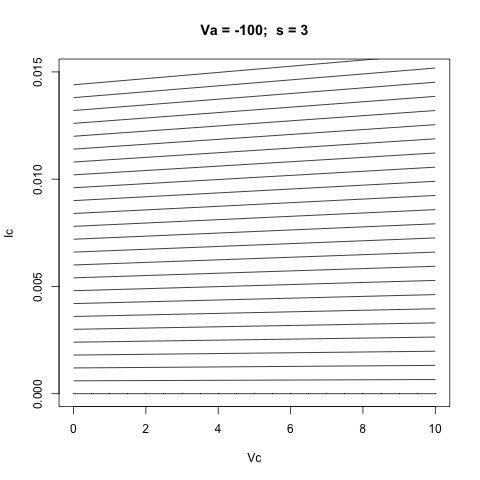} &   \includegraphics[width=6cm]{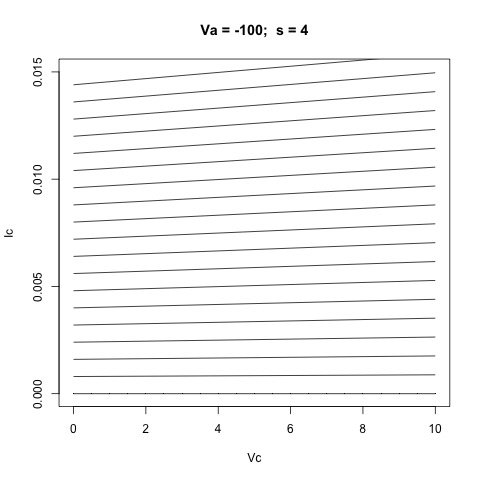} &   \includegraphics[width=6cm]{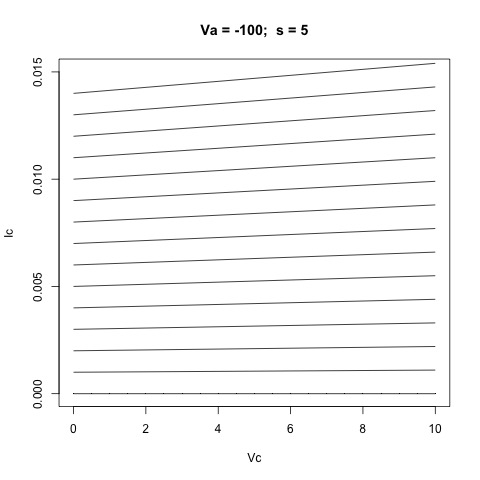}\\
    \end{tabular}
    \caption{Isolines defined by $I_b$ equispaced by $\Delta I_b = 2 \mu A$ in a region of the $V_C \times I_c$ space, for
                  several combinations of $V_a$ and $s$ values.  It does not necessarily follow that BJTs with larger $V_a$ 
                  will have larger or smaller average $\beta$. }
    \label{fig:earlycases}
\end{table*}

Because BJTs have been traditionally characterized by the $\beta$ and $R_o$ parameters, it is interesting to derive 
relationships between these parameters and  $V_a$ and $s$.   One of the disadvantages of the more traditional approach
is that $\beta$ and $R_o$ vary along the $V_C \times I_C$ characteristic space.  Indeed, both these parameters will
continuously change as the transistor voltages and currents vary during normal operation.  So, we need to consider the average
values of these two parameters within a given region (operation space), which are henceforth expressed as 
$ \langle \beta \rangle $ and $ \langle R_o \rangle$.   Other regions, or even curves (e.g. load lines), of interest can be taken into account
while deriving these averages.  
These relationships, derived analytically by integrating the two parameters along the operation region $Q$, are given
as follows:

\begin{table*}
\begin{eqnarray} \label{eq:rels}
	\langle \beta \rangle = \frac{1}{A} \int_{0}^{V_{C,max}} \int_{0}^{I_{C,max}}  \beta(I_C,V_C) dV_C dI_C =   \nonumber \\
   =  \frac{s}{6 V_{C,max}} \left[   2 I_{C,max}^2 ln \left( \frac{V_a-V_{C,max}}{V_a} \right) + 3 V_{C,max}^2 - 6 V_a V_{C,max}  \right] ,
         \label{eq:av_beta}\\
        \langle R_o \rangle =  \frac{1}{A} \int_{0}^{V_{C,max}} \int_{I_{C,min}}^{I_{C,max}}  R_o(I_C,V_C) dV_C dI_C
                  =  \frac{\left( V_{C,max}^2 - V_{C,max} V_a \right) }{V_{C,max}I_{C,max}} ln \left( \frac{I_{C,max}} {I_{C,min}} \right),  
         \label{eq:av_Ro}
\end{eqnarray}
\end{table*}  

where $A$ is the area of the region through which the parameters are integrated.  In the present case, it corresponds
to the area of the operation region $Q$, so that $A = (I_{C,max} - I_{C,min})/V_{C,max}$.

\begin{figure*}[h!]
\centering{
\includegraphics[width=14cm]{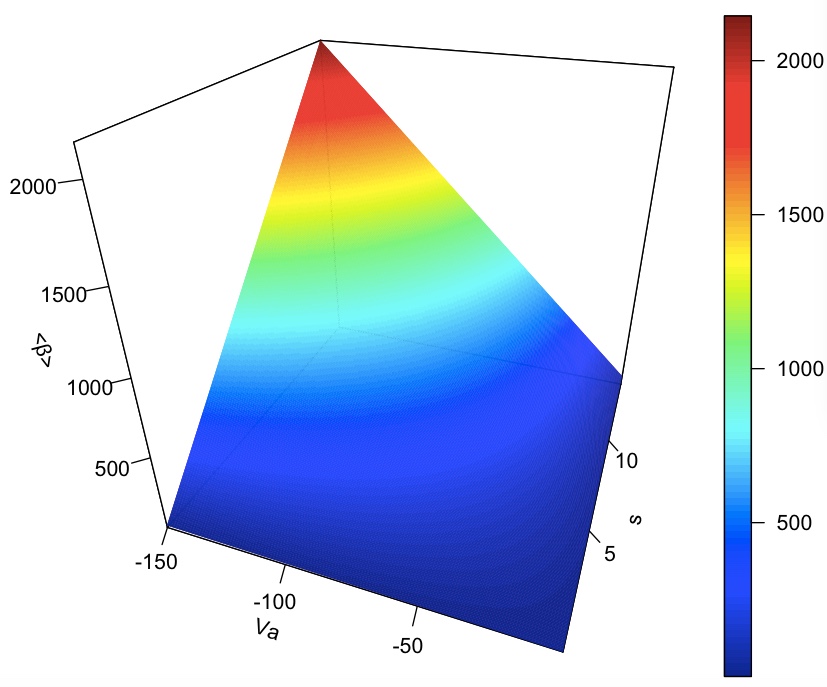}
\caption{The values of $\langle \beta \rangle$ defined by all the parameter configurations in the Early space 
region defined by $-150V \leq V_a \leq 0V$ and $0 \leq s \leq 14$.   The average current gain tends to increase
in a non-linear way with  $V_a$ magnitude and linearly with $s$, reaching a peak larger than 2000 at $(V_a = V_{a,max}, s = s_max)$.  }
\label{fig:Early3D}}
\end{figure*}

It follows from these relationships that $\langle \beta \rangle$ varies in a non-linear way with $V_a$ and linearly with $s$.  The 
output resistance, however, does not vary with $V_a$.  Figure~\ref{fig:Early3D} depicts the values of $\langle \beta \rangle$
in terms of $V_a$ and $s$, for $-150V \leq V_a \leq 0V$ and $0 \leq s \leq 14$, respectively.  A peak average current gain
exceeding 2000 is observed, corresponding to the parameter configuration $(V_a = V_{a,max}, s = s_{max})$.  Because typical 
small signal BJTs are known to have average current gain nearly between 50 and 500, only a relatively small region of the 
$V_a \ times s$ space will be populated by real-world devices, which is indeed confirmed in the experimental part of this work.

\begin{figure}[h!]
\centering{
\includegraphics[width=8cm]{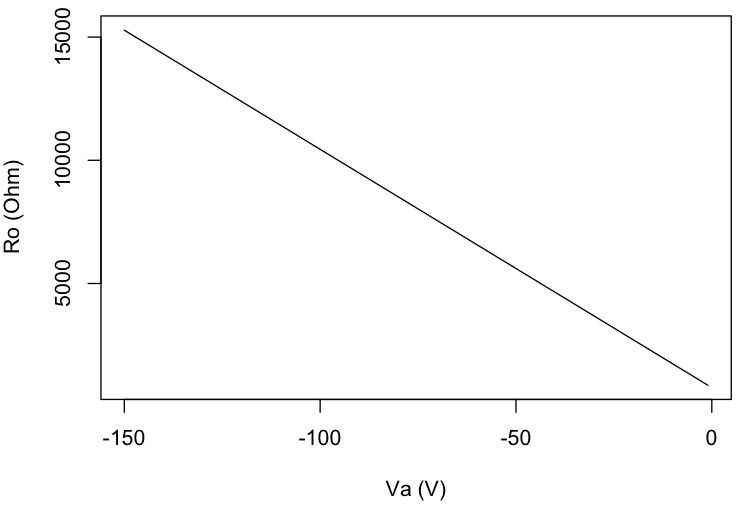}
\caption{Linear variation of $R_o$ with the values of $V_a$ obtained for the considered BJTs.   }
\label{fig:Ro}}
\end{figure}

The linear relationship between $\langle R_o \rangle$ and $V_a$ is presented in Figure~\ref{fig:Ro}.  It follows
that increased output resistances will be obtained for large magnitude values of $V_a$. 

\section{Numerical Estimation of Early Model Parameters}

So far we have discussed the characteristics and implications of the Early modeling of BJTs on a purely analytical
basis.  In order to extend the Early-based analysis to real-world devices, so as to better understand the statistical properties
of the transistors parameters and their interrelationship, it is necessary to resource to experimental approaches, which
constitutes one of the objectives of the current work.   More specifically, we need the means for, given a set of real-world
devices, deriving their isolines (level-sets) and estimating the respective Early parameters $V_a$ and $s$, as well as the more
traditional $\langle \beta \rangle$ and $\langle R_o \rangle$.  We have already discussed in Section~\ref{sec:acq} how 
the adopted acquisition system is capable of obtaining the voltages $V_{CC}$, $V_C$, $V_{BB}$ and $V_B$ and the 
currents $I_C$ and $I_B$, as well as the pre-processing that is applied in order to obtain angularly equispaced isolines and 
reduce acquisition and other types of noise.  These isolines can now be numerically handled in order to estimate the Early parameters.   
An experimental approach to Early modeling has been described~\cite{costaearly:2017} (see also~\cite{costafeed:2017} for the 
estimation of Early voltage) which involves linear regression of 
the isolines and respective identification of the intersections of prolongations until crossing the $V_C$ axis by considering
averages or peaks of the density of crossings along the $V_a$ axis.   Though that 
approach has provided several useful results, here we adopt a different procedure intended to minimize the 
interference between the isoline prolongations as would be the case while taking the average of the intersection voltage 
values in order to estimate $V_a$.   Instead, we use an adaptation of the voting (or accumulating) principle which has 
traditionally been one of the stages of the Hough transform for detection of straight lines 
(e.g.~\cite{hough:1962,shapebook,costa:1993}).  The Hough transform is an image analysis method capable of identifying 
the presence of multiple lines (or other types of curves) in an image even in presence of noisy and occlusion.  This 
method, when applied for straight lines, involves mapping each of the foreground image pixels into 
a curve (e.g. a sinusoidal or straight line) in a respective two-dimensional parameter space, discretely represented as 
the \emph{accumulator array}.  Each point of this array covered by one of the mapping curves is incremented (voting), so 
that peaks in the accumulator array indicate the putative presence of respective straight lines in the image, with the 
coordinates of these peaks corresponding to the respective straight line parameters.  This accumulating, or \emph{voting}, 
principle is adopted here in order to identify intersections between isolines among themselves, and not necessarily 
with the $V_C$ axis.  It has been experimentally verified that such intersections sometimes appear in the case 
of real-world BJTs.  In addition to generalizing the identification of intersections out of
the $V_C$ axis, this accumulating approach also allows less interference between the isolines, as would be
otherwise obtained by taking averages of the intersections with the $V_C$ axis.  Another advantage of this approach
is that it becomes less critical and easier to restrict the isolines in order to avoid the saturation and cut-off effects on the 
straight portions of these curves.

\begin{figure}[h!]
   \includegraphics[width=7cm]{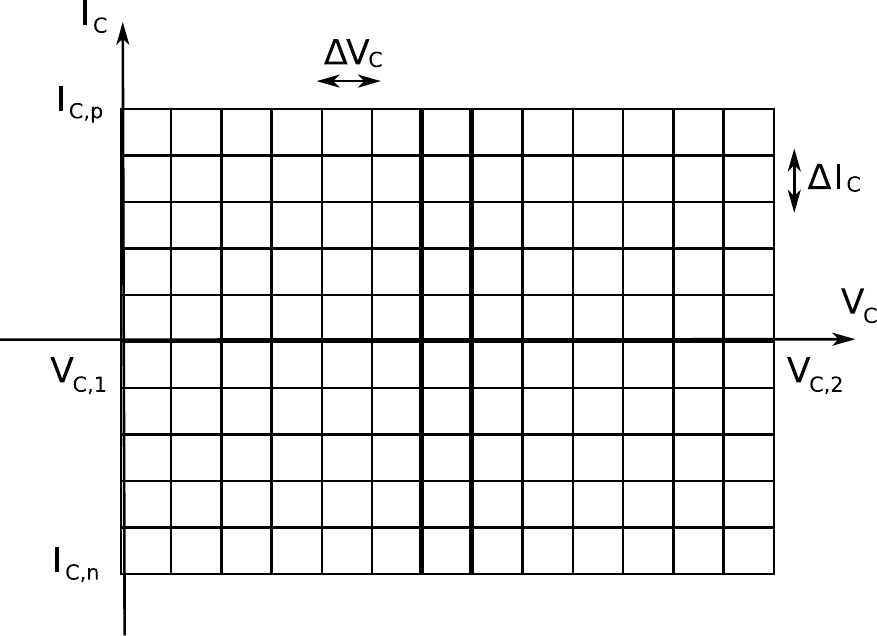}
   \caption{The accumulator array used for $V_a$ numerical estimation, indexed as $Acc[v_C, i_C]$, with resolutions
           $\Delta V_C$ and $\Delta I_C$.}
   \label{fig:accum}
\end{figure}

The Hough transform-inspired method to detect $V_a$ and $s$ is presented as follows.  First, we adopt straight 
$(m,c)$ parameterization, instead of the polar $(\alpha, \rho)$ parameterization commonly used in the Hough transform.
This parameterization of straight lines is suitable for our purposes here, because the straight lines to be detected
(the prolongations of the straight portions of the isolines) can be geometrically normalized so as to keep the magnitude
of the slope parameter $m$ small (problems are implied when the slope reaches 1).  The accumulator array is chosen to 
be nearly square (in 
the sense of having the comparable vertical and horizontal divisions) in order to allow for better separation between 
the straight lines.  Figure~\ref{fig:accum} illustrates the accumulator structure $Acc[v_C, i_C]$, corresponding to 
the parameter space region defined by $V_{C,1} \leq V_C \leq V_{C,2}$ and 
 $I_{C,n} \leq V_{C} \leq I_{C,p}$, with $I_{C,n} = - I_{C,p}$ for ensuring symmetry.  We henceforth take
 $V_{C,1} = -150V$, $V_{C,2} = 0V$, $I_{C,p} = I_{C,max}/4 \approx 5mA$  The resolution along the $V_C$ and
 $I_C$ axes are identified as $\Delta V_C$ and $\Delta I_C$, respectively.

Each isoline $i$ identified as described in Section~\ref{sec:acq} has its parameters represented as $(m_i,c_i)$.  So, the
accumulation can proceed by calculating, for $i_{C,1} \leq i_{C,2}$,  the value $i_c(v_c)$  as:

\begin{equation}
   i_c(v_C) = round((m_i~v_C + c_i-I_{C,n}) / \Delta I_C).
\end{equation}

The respective accumulator cell $Acc[i_C, v_C]$ is incremented for each obtained pair $(i_C,v_C)$. 
Several methods have been proposed (e.g.~\cite{costa:1989,costa:1990,shapebook}) in order to improve the voting 
scheme, such as updating also neighboring accumulator cells around $Acc[i_C, v_C]$ and using interval arithmetics.
Thus, each intersection between two or more isoline prolongations will result in a count value larger than 1.  Once the
accumulating procedure is complete, we seek for the maximum count in the accumulator and takes the
respective coordinate $V_{C,peak}$ as an estimation of $V_a$.   Figure~\ref{fig:exmpl_acc}
illustrates an example of the procedure considering the isolines obtained for a real-world BJT, with $V_C$ in Volts
and $I_a$ in $mA$.  The prolongations of the isolines of this real-world BJT converge at a point $V_C,I_C$ 
which is emphasized in white.  Observe that, as observed above, sometimes the convergence point results with 
$I_C \neq 0$, but this value has been found to be typically very small, in the order of $0.1mA$.

\section{Results and Discussion}

This section reports and discusses the main experimental results obtained in this article.  We henceforth assume
$V_{C,max} = 8V$, $I_{C,min} = 1mA$, $I_{C,max} = 15mA$, and $N_i = 100$ angularly equispaced isolines.

\subsection{Experimental BJTs}

The experiments reported in this work employied 7 BJT complementary pairs,
each of them represented as $\pi_i = (N_i, P_i)$, $i = 1, \ldots, 7$, respectively to the
NPN and PNP cases.  These devices were chosen as they are frequently used in practice.  
Ten samples of each transistor type were considered, all from the same
lot.  Pairs $\pi_1$ to $\pi_4$ are members of a same family, as well as $p_5$ and $p_6$.
Table~\ref{tab:feats} gives some of the main electronic features of the chosen BJT pairs.

\begin{table*}
  \centering
     \begin{tabular}{| c || c | c | c | c | c | c | c | c | c | c}
        \hline
        BJT pair & $\langle \beta \rangle$ & $I_{C,max}~(mA)$  & $V_{CEO}~(V)$ &  $f_T~(MHz)$ & relative features  \\
        \hline \hline
        $\pi_1$ & 300 & 100 & 60 & 150 &  large $V_{CEO}$ \\ 
        $\pi_2$ & 450 & 100 & 40 & 150 &   -- \\ 
        $\pi_3$ & 300 & 500 & 30 & 300 &  large $f_T$ \\ 
        $\pi_4$ & 300 & 100 & 30 & 250 &   linearity, low noise  \\ 
        $\pi_5$ & 350 & 800 & 40 & 200 &   large $I_{C,max}$  \\ 
        $\pi_6$ & 350 & 800 & 20 & 200 &   large $I_{C,max}$  \\ 
        $\pi_7$ & 200 & 200 & 40 &  65 &   -- \\ 
        \hline
      \end{tabular}
    \caption{Typically expected electronic features of the BJT complementary pairs used in the reported experiments.}
    \label{tab:feats}
\end{table*}

The signals of these devices were obtained by the acquisition system, and the above described
procedures were applied to each of them.  The experiments were performed for
$0V \leq V_{CC} \leq 12V$ and $0V \leq V_{CC} \leq 2V$ for the NPN
devices, and for $0V \leq V_{CC} \leq  12V$ and $0V \leq V_{BB} \leq  2V$ 
otherwise.  A total of 100 angularly equispaced isolines were obtained for each device out of the
original 256 values of $V_{BB}$, while $V_{CC}$ was sampled with 128 values.  The last
80 values of each isoline were used for obtaining the isolines prolongations onto
the $V_C$ axis, which is done by minimum squares linear regression (e.g.~\cite{shapebook}).  The Hough 
voting scheme was performed with $V_{C,1} = -150V$
$V_{C,2} = 0$,  $\Delta V_C = 1$, $I_{C,n} = -5mA$, $I_{C,p} = 5 mA$, 
and $\Delta I_C = 500 \mu A$.  The voting scheme was applied not only for each generated 
point $(i_C, v_C)$, but also for the 8 nearest neighboring cells, as this procedure
allows additional tolerance to eventual noise in the isolines.

\begin{figure}[h!]
\centering{
\includegraphics[width=8cm]{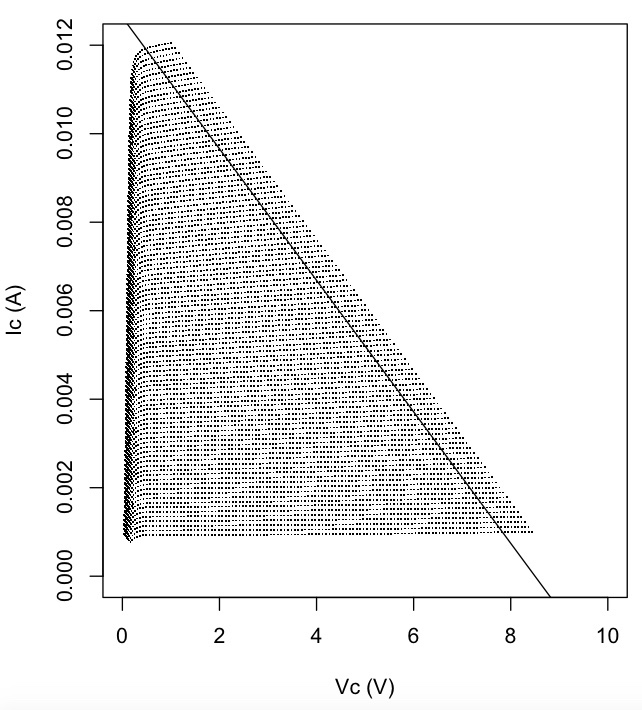}
\caption{The isolines obtained for BJT ($N_6(7)$).  The load line used for signal capture is
also included.}
\label{fig:exmpl_iso} }
\end{figure}

We now proceed to illustrate the main data and results, as derived for one of the considered BJTs ($N_6(7)$).
Figure~\ref{fig:exmpl_iso} shows the angularly equispaced isolines obtained for this device,
together with a load line obtained for $V_C = 8.5V$ and the resistance $R_L = 673 \Omega$, the latter being used
for scanning the device voltages and currents.  The saturation region can be identified at the
righthand side of the figure.

\begin{figure}[h!]
\centering{
\includegraphics[width=8cm]{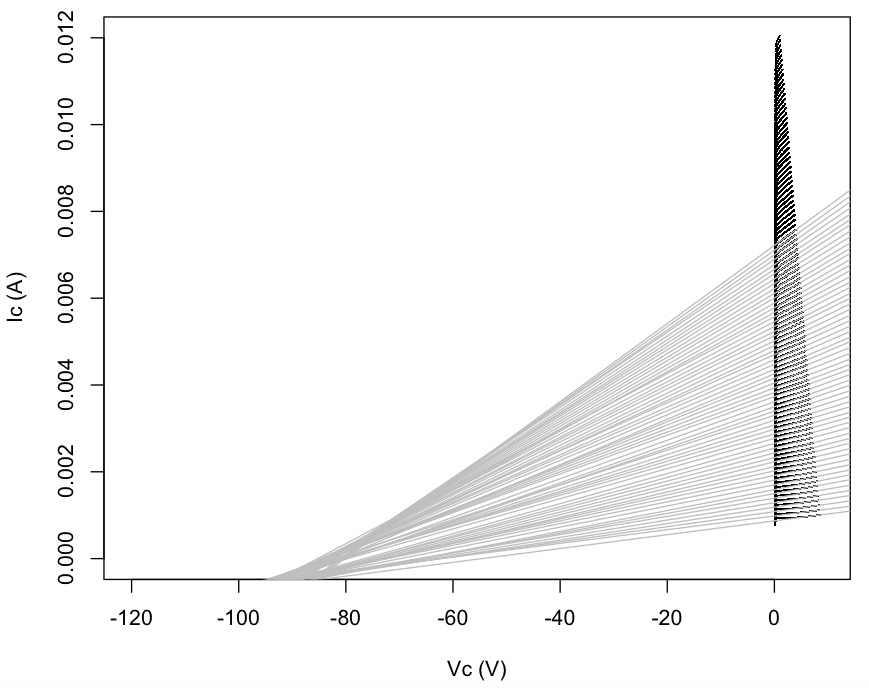}
\caption{The isoline prolongations obtained by linear regression of the isolines for the BJT ($N_6(7)$).}
\label{fig:exmpl_prolong} }
\end{figure}

The obtained prolongations of the isolines, shown in Figure~\ref{fig:exmpl_prolong}, tend to
intersect at a well-defined point close to the $V_c$ axis.  

\begin{figure*}[h!]
\centering{
\includegraphics[width=15cm]{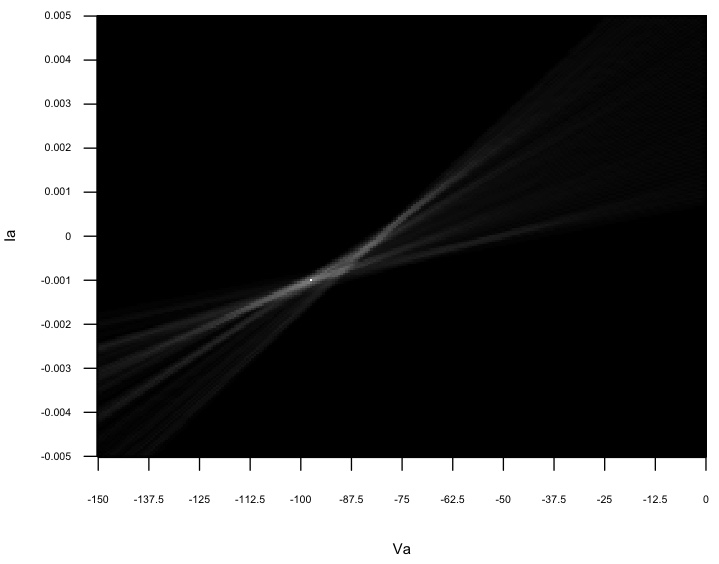}
\caption{Example of accumulator array typically obtained for the considered devices.  }
\label{fig:exmpl_acc} }
\end{figure*}

Figure~\ref{fig:exmpl_acc} illustrates the accumulator array obtained for the considered
device.   Observe that the isolines tend to form groups, such as that emanating from the
upper portion of the characteristic surface.   The adopted voting scheme allows the identification
of the most likely intersection, reducing the interference between the diverging isolines, which
often are obtained for the most extreme isolines in the characteristic space.  
Observe also that the resulting accumulation peak has non-null, though very small, $I_C$
value, which is henceforth denominated the \emph{Early current} $I_a$.   

\begin{figure}[h!]
\centering{
\includegraphics[width=6cm]{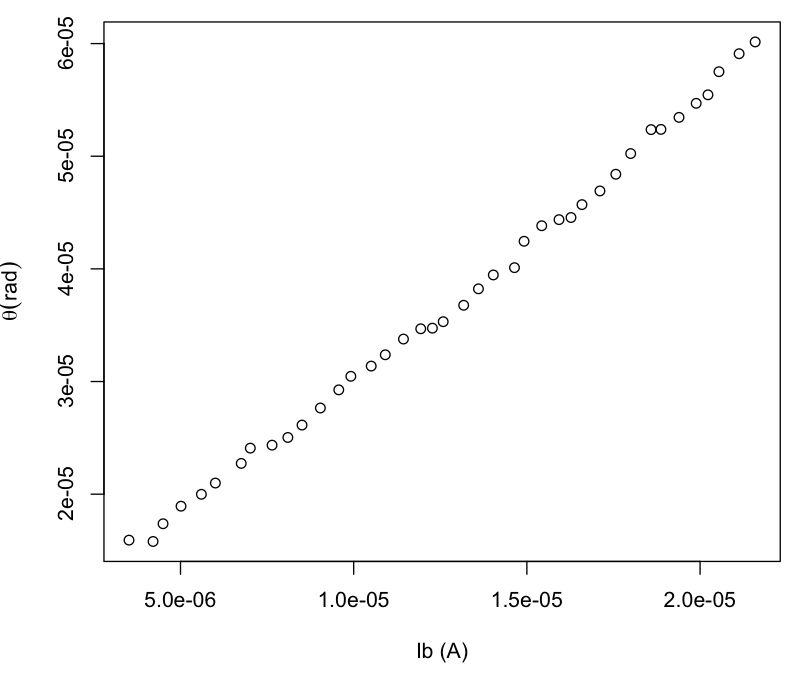}
\caption{The well-defined linear relationship between $\theta$ and $s$ observed for all considered
devices, including the BJT ($N_6(7)$) in this example.}
\label{fig:exmpl_lin} }
\end{figure}

The linear relationship between $\theta$ and $s$ generally observed for the considered devices are 
shown in Figure~\ref{fig:exmpl_lin}.  The proportionality parameter $s$ can be estimated by performing linear
regression on this scatterplot, yielding also the intersection value of the line with the $\theta$ axis,
which is henceforth denoted as $cs$.

\begin{figure}[h!]
\centering{
\includegraphics[width=8cm]{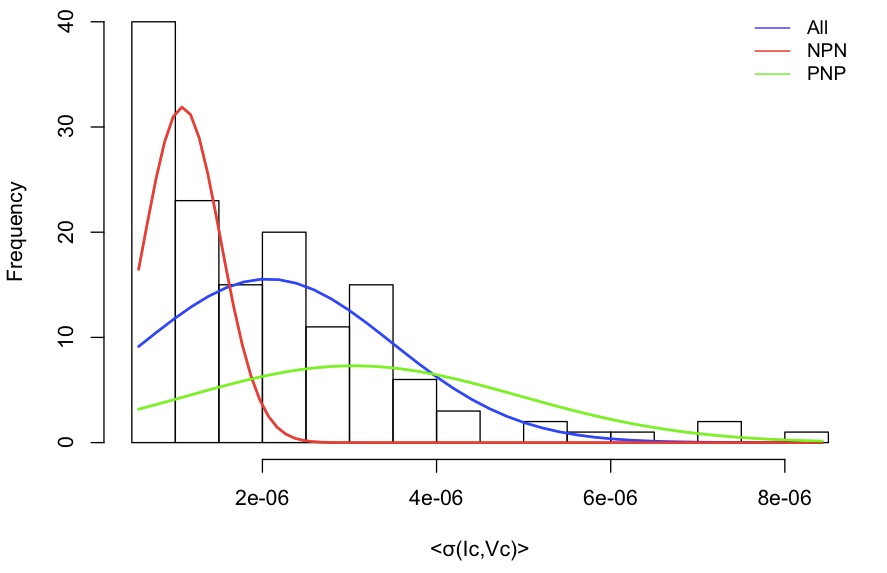}
\caption{The density of the residual standard deviation sigma obtained for estimating the relationship between
$V_C$ and $I_C$ (isoline prolongations).}
\label{fig:IcVc} }
\end{figure}

\begin{figure}[h!]
\centering{
\includegraphics[width=8cm]{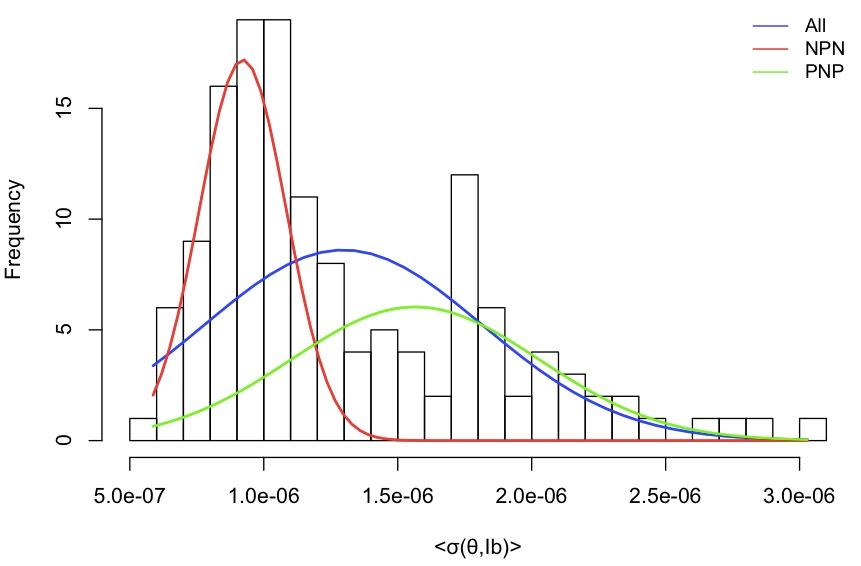}
\caption{The density of the residual standard deviation sigma obtained for estimating the relationship between 
$\theta$ and $Ib$.}
\label{fig:thIb} }
\end{figure}

The accuracy of the estimation of the Early parameters was quantified in terms of the
residual standard deviation  sigma for each case.  Figure~\ref{fig:IcVc} shows the
sigma for estimating the isoline prolongations (obtained by linear regression of the
respective isolines), and~\ref{fig:thIb} gives the 
sigma for estimating the relationship between $\theta$ and $I_B$.  In both these figures,
normal fitting s are included with respect to all the BJTs as well as the two groups NPN and PNP.
The obtained values are small and corroborate the accuracy
of the estimations.  Interestingly, larger errors are typically obtained  for the PNP transistors, as these
tend to present larger parameter variability.

\section{BJT Individual Analysis}

Tables~\ref{tab:Vas} presents the $V_a$ values obtained for each of the 140 considered
BJTs, and table~\ref{tab:ss} gives the respective $s$ values.

\begin{table}
  \centering
     \begin{tabular}{c || r r r r r r r r r r}
        \hline
        BJT & 1 & 2 & 3 & 4 & 5 & 6 & 7 & 8 & 9 & 10 \\
        \hline \hline
        N1 & -76 & -93 & -91 & -90 & -72 & -71 & -83 & -86 & -83 & -78 \\ 
        P1 & -39 & -35 & -38 & -39 & -38 & -58 & -46 & -48 & -43 & -45 \\ 
        N2 & -99 & -91 & -84 & -95 & -90 & -94 & -90 & -95 & -99 & -89 \\ 
        P2 & -62 & -50 & -37 & -40 & -41 & -21 & -56 & -50 & -38 & -57 \\ 
        N3 & -94 & -97 & -95 & -98 & -96 & -94 & -96 & -89 & -91 & -95 \\ 
        P3 & -57 & -48 & -59 & -52 & -35 & -54 & -20 & -60 & -55 & -57 \\ 
        N4 & -125 & -86 & -78 & -101 & -96 & -104 & -79 & -73 & -68 & -94 \\ 
        P4 & -36 & -30 & -40 & -43 & -42 & -42 & -46 & -43 & -39 & -39 \\ 
        N5 & -100 & -102 & -112 & -103 & -113 & -93 & -109 & -102 & -109 & -112 \\ 
        P5 & -61 & -58 & -68 & -79 & -61 & -63 & -61 & -73 & -66 & -63 \\ 
        N6 & -96 & -90 & -92 & -90 & -98 & -97 & -92 & -91 & -99 & -96 \\ 
        P6 & -51 & -22 & -48 & -23 & -37 & -38 & -29 & -34 & -36 & -35 \\ 
        N7 & -114 & -120 & -115 & -109 & -131 & -109 & -107 & -136 & -110 & -104 \\ 
        P7 & -25 & -37 & -48 & -46 & -60 & -29 & -36 & -24 & -32 & -28 \\
        \hline
      \end{tabular}
    \caption{The $V_a$ values obtained for each of the 140 considered BJTs,
                   organized as complementary pairs.}
    \label{tab:Vas}
\end{table}

\begin{table}
  \centering
     \begin{tabular}{c || r r r r r r r r r r}
        \hline
        BJT & 1 & 2 & 3 & 4 & 5 & 6 & 7 & 8 & 9 & 10 \\
        \hline \hline
      N1 & 3.67 & 3.17 & 3.22 & 3.26 & 3.31 & 4.16 & 3.21 & 3.32 & 3.44 & 3.56 \\ 
      P1 & 4.68 & 5.52 & 6.58 & 4.89 & 7.1 & 3.08 & 3.36 & 4.15 & 4.59 & 5.56 \\ 
      N2 & 2.47 & 2.60 & 2.86 & 2.42 & 2.68 & 2.58 & 2.7 & 2.56 & 2.52 & 2.67 \\ 
      P2 & 3.43 & 4.27 & 6.48 & 5.59 & 4.9 & 14.46 & 3.68 & 4.27 & 6.01 & 3.67 \\ 
      N3 & 2.54 & 2.46 & 2.58 & 2.48 & 2.56 & 2.61 & 2.54 & 2.71 & 2.67 & 2.53 \\ 
      P3 & 3.33 & 4.34 & 2.84 & 3.61 & 4.65 & 3.45 & 14.01 & 2.73 & 3.38 & 3.17 \\ 
      N4 & 0.85 & 2.15 & 3.28 & 1.96 & 2.19 & 2.04 & 3.41 & 4.2 & 3.52 & 2.08 \\ 
      P4 & 6.67 & 8.18 & 6.02 & 4.97 & 5.97 & 5.83 & 5.28 & 5.73 & 6.24 & 6.11 \\ 
      N5 & 2.24 & 2.56 & 2.55 & 2.62 & 2.29 & 2.95 & 2.51 & 2.55 & 2.56 & 2.59 \\ 
      P5 & 3.78 & 4.15 & 2.57 & 2.60 & 3.87 & 3.35 & 4.05 & 2.96 & 3.34 & 3.00 \\ 
      N6 & 2.32 & 3.80 & 2.53 & 2.42 & 2.27 & 2.39 & 2.45 & 2.54 & 2.33 & 2.45 \\
      P6 & 3.65 & 12.8 & 2.39 & 11.32 & 6.16 & 5.18 & 8.97 & 8.07 & 6.69 & 7.11 \\ 
      N7 & 1.24 & 1.35 & 1.43 & 1.67 & 1.13 & 1.62 & 1.82 & 1.05 & 1.12 & 1.34 \\ 
      P7 & 6.29 & 3.60 & 2.75 & 3.00 & 2.34 & 6.01 & 3.34 & 7.27 & 5.13 & 5.89 \\ 
      \hline
      \end{tabular}
    \caption{The $s$ values obtained for each of the 140 considered BJTs,
                   organized as complementary pairs.}
    \label{tab:ss}
\end{table}

\begin{figure*}[h!]
   \subcaptionbox{$V_a$\label{fig1:a}}{\includegraphics[width=8cm]{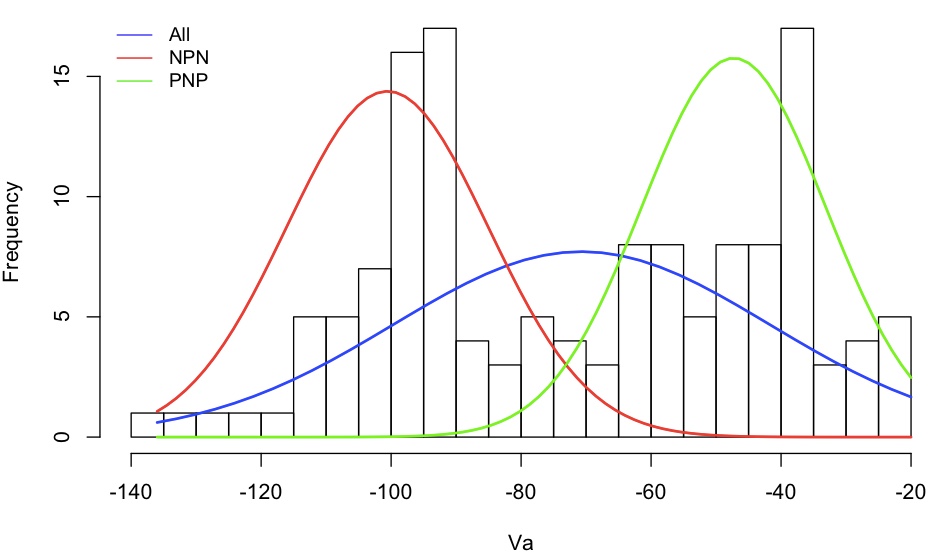}} \hspace{1cm}
   \subcaptionbox{$s$\label{fig1:b}}{\includegraphics[width=8cm]{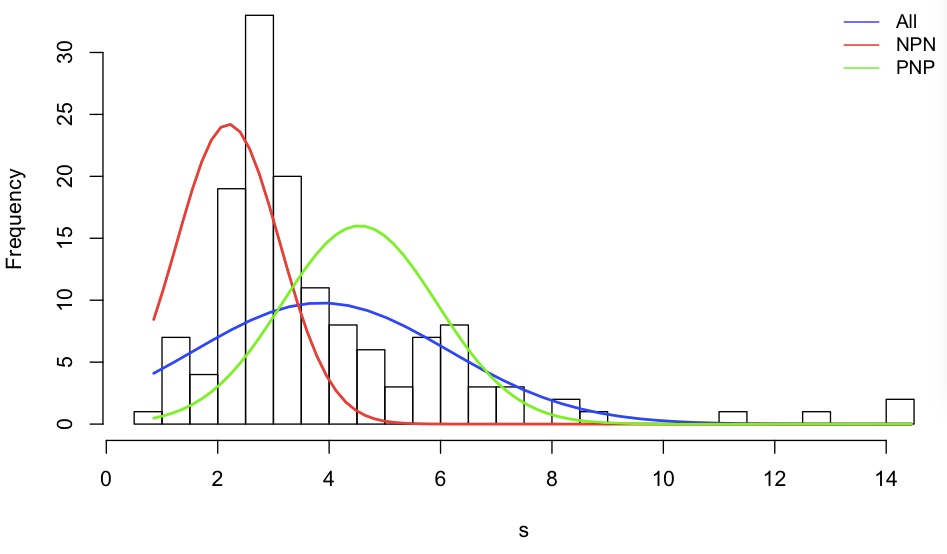}}  \\ \vspace{1cm}
   \subcaptionbox{$I_a$\label{fig1:a}}{\includegraphics[width=8cm]{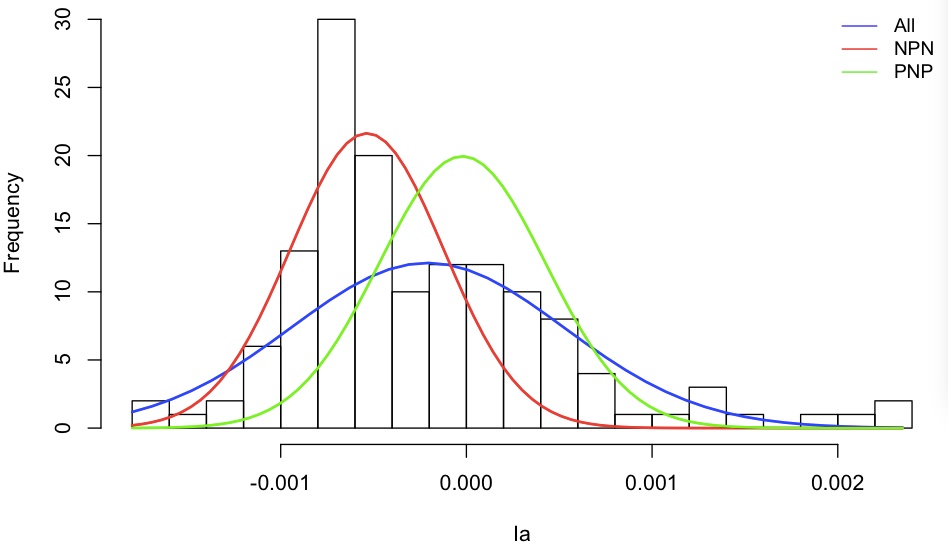}} \hspace{1cm}
   \subcaptionbox{$cs$\label{fig1:b}}{\includegraphics[width=8cm]{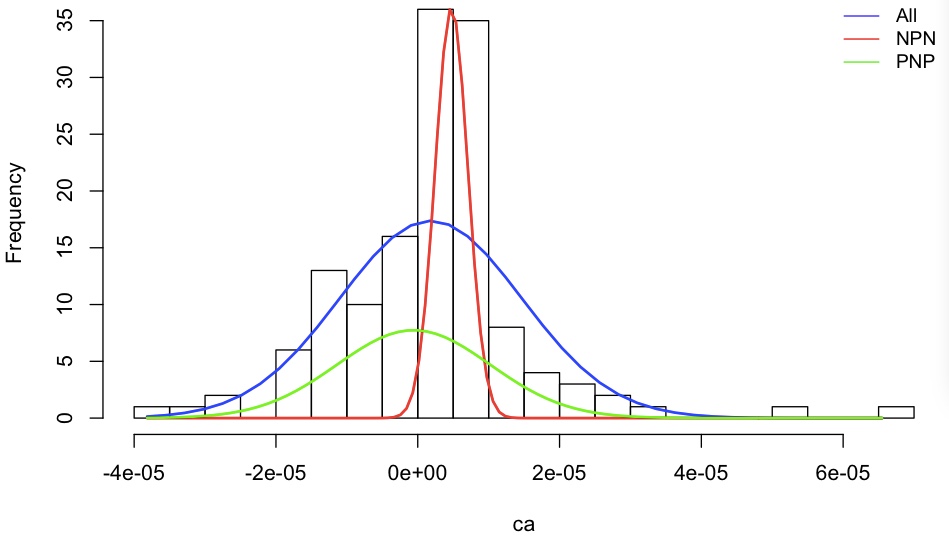}} \\ \vspace{1cm}
   \subcaptionbox{$acc$\label{fig1:a}} {\includegraphics[width=8cm]{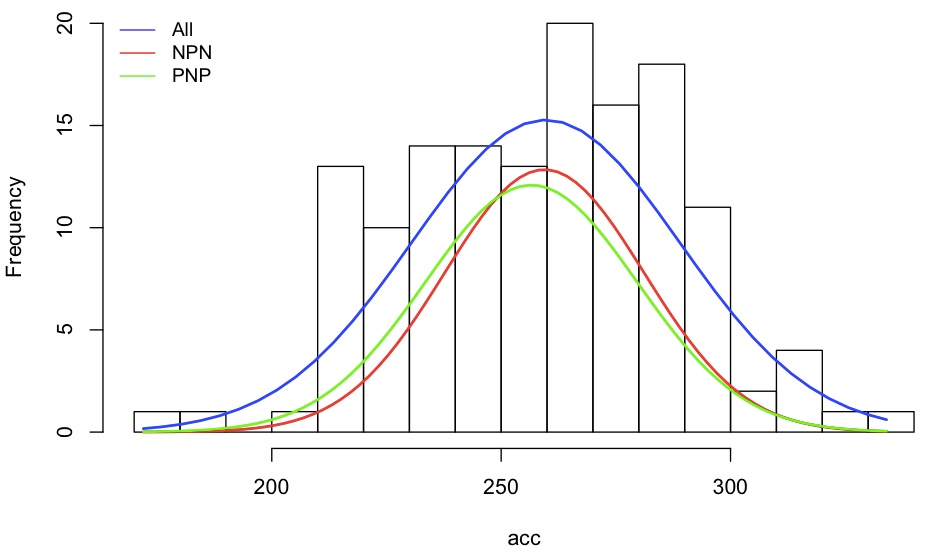}} 
 \caption{Histograms of several of the parameters associated with the Early model, with respect to the 140 
 adopted real-world BJTs, with superimposed normal fittings.}
\label{fig:hists}
\end{figure*}

Each of the considered parameters (i.e. $V_a$, $s$, $I_a$, $ca$, $acc$) had their histograms (densities) estimated
considering all pairs and by NPN or PNP type.  The results are presented in Figure~\ref{fig:hists}.   These
densities provide valuabl information about the behavior and parameter variability of the considered BJTs.

In Figure~\ref{fig:hists}(a), it is showed the distribution of the estimated Early voltages $V_a$.  It follows from this
result that the two groups of transistors, NPN and PNP, are markedly distinct one another with respect to this
parameter, with the latter devices being characterized by smaller magnitude of $V_a$, implying in smaller
output resistance $R_o$ but also less linearity and possibly smaller gain than the NPN transistors.  This observed
tendency is probably related to the fact that NPN are typically employed in practice more often than PNP
devices. 

The distribution of the values of the Early parameter $s$ obtained for the chosen BJTs are shown in Figure~\ref{fig:hists}(b).
Differently from what was observed for $V_a$, the distribution of $s$ is considerably narrower for the NPN transistors,
suggesting that this type of devices has an inherently smaller parameter variability.  Overall, BJTs tend to have their $s$
parameter values near 3, where the density peaks.

The parameter that we have called Early current, namely $Ia$, has its distribution shown in Figure~\ref{fig:hists}(c).  This
density function demonstrates that this parameter has very small values, with a well-defined peak near $I_a = -0.0008A$.
These results suggest that PNP transistors tend to have larger $I_a$ values.  

The distribution of $cs$ values, pertaining to the intersection parameter obtained in the linear regression used for
estimating the relationship between $\theta$ and $|I_B$ are presented in Figure~\ref{fig:hists}(d).  The NPN devices
has a very sharp peak near $cs=0$, indicating almost no offset in the relationship $\theta = s I_B$ observed for the 
considered BJTs.  A much wider distribution is verified for the PNP devices, possibly as a consequence of their
larger parameter variability.

Figure~\ref{fig:hists}(e) presents the accumulation values obtained while estimating each of the pairs of parameter
$(V_a,s)$ for the 140 considered transistors.  Very similar densities were obtained for both NPN and PNP, with
the most common number of counts corresponding to 250 votes, which corroborates the stability of the method.

\subsection{Relationships among Early Parameters}

We now proceed to analyzing the obtained results in pairwise fashion, so as to better characterize parameter
variability and to identify possible relationships.

\begin{figure*}[h!]
\centering{
\includegraphics[width=15cm]{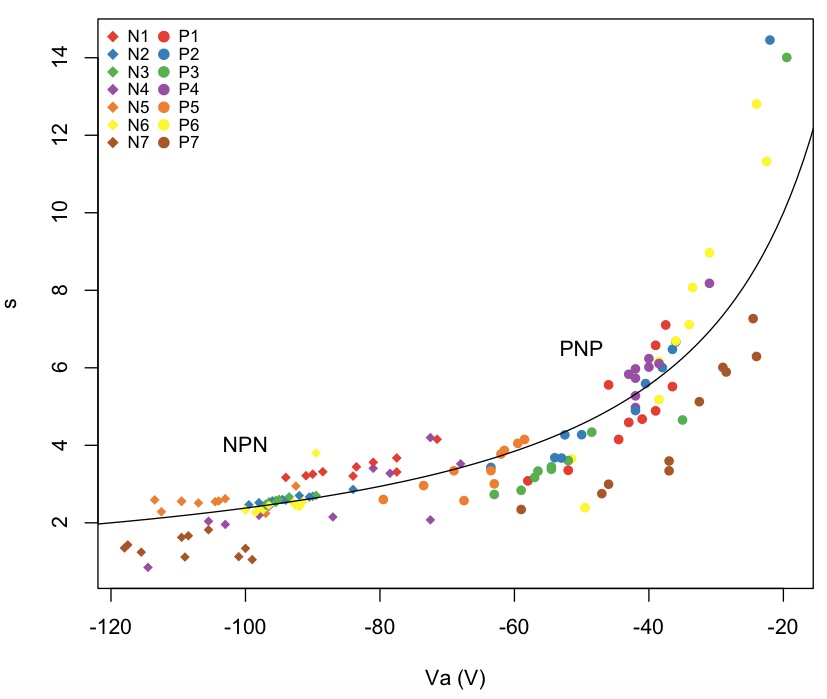}
\caption{Scatterplot of $V_a$ in terms of $s$ parameter obtained for the considered BJTs.   An impressive separation
between the NPN and PNP groups is immediately observed, with the former occupying the lefthand portion of the plot.
A well-defined relationship is also verified, with $s$ increasingly slowly with $V_a$ up to about $V_a = -45V$, undergoing
a steeper increase afterwards. Regarding the relative distance between respective NPN and PNP devices of the same 
complementary pair, the nearest case is observed for $\pi_5$, and the furthest for $\pi_7$.  Remarkably, the NPN group
presents intrinsic smaller parameter dispersion (variability) than the PNP group. The isoline obtained for $\langle \beta 
\rangle = 250$ is also included, corresponding to a kind of medial axis to the curved band occupied by the devices.  }
\label{fig:VaXs} }
\end{figure*}

Figure~\ref{fig:VaXs}, which is probably one the most important contributions of the present work.
shows the scatterplot defined by the Early voltage $V_a$ and proportionality parameter $s$ obtained for all the 
considered BJTs.  Several remarkable results can be identified.  First, we have that all the considered BJTs occupy
a relatively narrow $V-$like strip along the Early space.  This region progresses from smaller values of $s$ of about
2, verified for the NPN transistors, to ever increasing $s$ parameters observed for the PNP devices at larger values of 
$V_a$.  Second, the two main types of transistors, NPN and PNP, are almost perfectly segregated one another,
with the former exhibiting larger magnitude of $V_a$ and smaller $s$, which favors linearity but also implies
larger output resistance. An opposite trend is observed for the PNP transistors.  In addition, because the latter type
has broader variation of $s$ and similar range of $V_a$, this group is characterized by larger parameter variability.
Observe the intense overlap between several of the considered NPN devices near the center of the respective cluster.

Interestingly, the curve narrow band defined by the isolines in the Early space and obeyed by the real-world
BJTs are remindful of a similar pattern found in a linear discriminant analysis (LDA) of traditional parameters of
arrays of NPN transistors~\cite{costaarray:2017}.  If that is so, it would corroborate further the usefulness
of using pattern recognition approaches to study semiconductor devices and circuits, which allowed that
prediction of the interesting relationship between the Early parameters.

\begin{figure}[h!]
\centering{
\includegraphics[width=8cm]{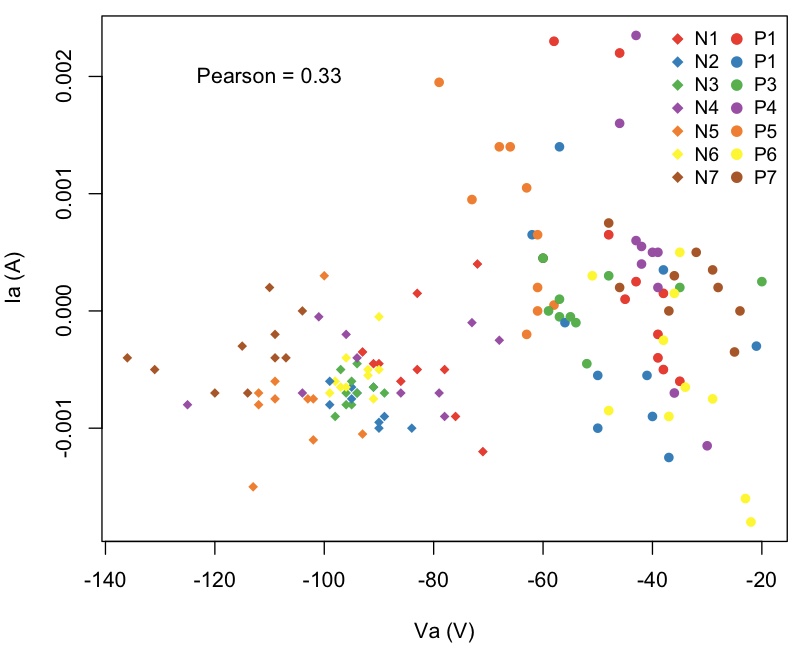}
\caption{Scatterplot of $V_a$ in terms of $I_a$ parameter obtained for the considered BJTs.   }
\label{fig:VaXIa} }
\end{figure}

Figure~\ref{fig:VaXIa} shows the pairwise relationship between $V_a$ and $I_a$.  At least for the
considered BJTs, no discernible relationship between these two parameters are observed, with a small 
Pearson cross-correlation of just 0.33.  Observe that the NPN and PNP groups are well-separated in this
scatterplot, exhibiting a more circular distribution.

\begin{figure}[h!]
\centering{
\includegraphics[width=8cm]{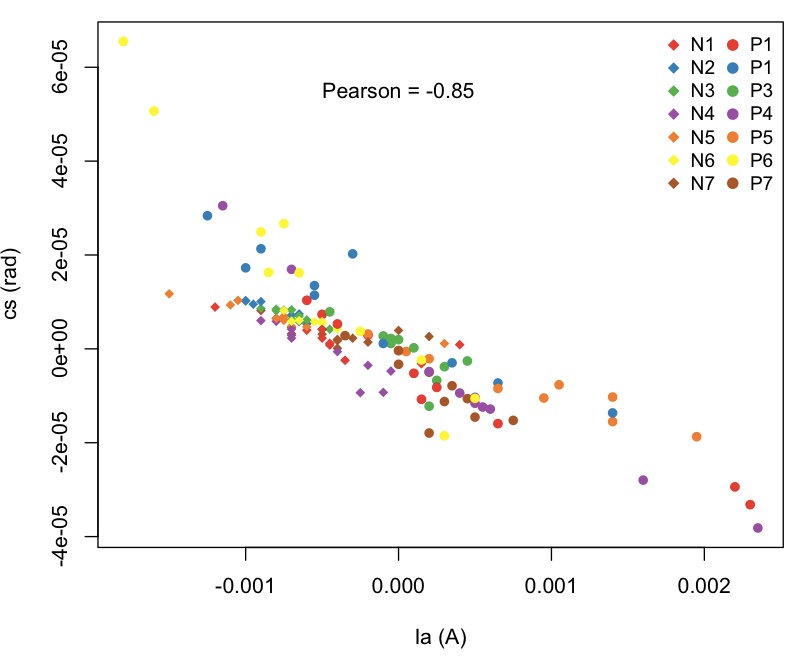}
\caption{Scatterplot of $I_a$ in terms of the $cs$ parameter obtained for the considered BJTs.  A strong negative
correlation is observed between these two measurements for the considered devices.}
\label{fig:IaXcs} }
\end{figure}

The relationship between the parameters 
$I_a$ and $cs$, shown in Figure~\ref{fig:IaXcs}, presented a strong negative correlation. This
correlation indicates that offset in the relationship between $\theta$ and $I_B$ (i.e. $\theta = s I_B + cs$) tends
to decrease in magnitude as $I_a$ increases.  Further investigation is required in order to better understand
this effect.

\begin{figure*}[h!]
\centering{
\includegraphics[width=15cm]{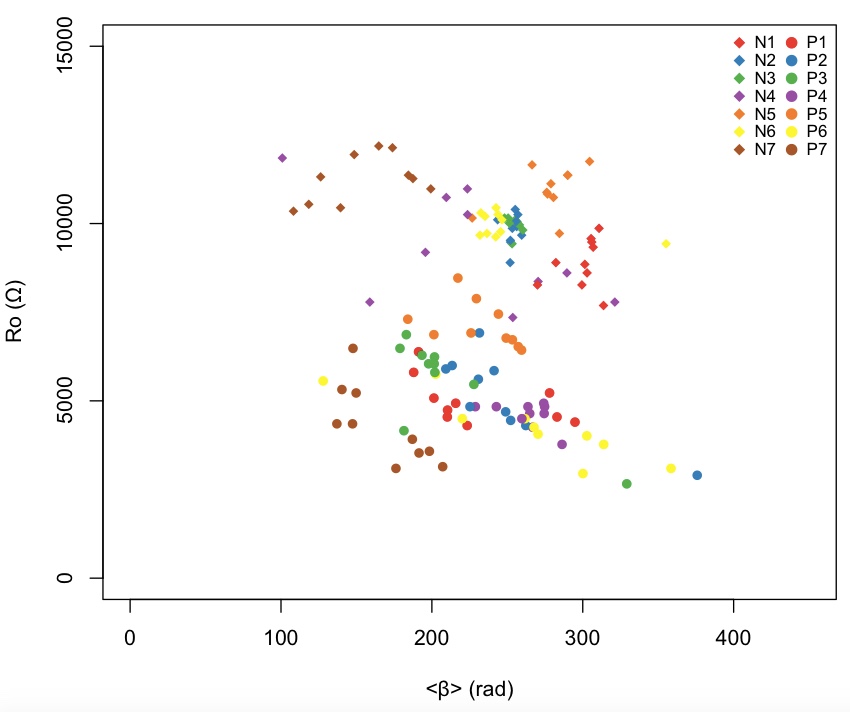}
\caption{Scatterplot defined by the $\langle \beta \rangle$ and $\langle R_o \rangle$ for the 140
considered BJTs.  The NPN and PNP also exhibit almost no overlap in this space, and the PNP devices
still exhibit, though to a smaller extent, larger parameter variation.  The 10 samples respective to 
each of the considered BJT types yielded well-defined respective clusters.  NPN devices have
larger $\langle R_o \rangle$ than the PNP counterparts, but the distribution of the values of $\langle \beta \rangle$
are mostly comparable.}
\label{fig:beta_Ro} }
\end{figure*}

Equations~\ref{eq:av_beta} and~\ref{eq:av_Ro} can now be applied to transform the $V_a \times s$ scatterplot 
obtained for the real-world transistors into the more traditional and electronically related $\langle beta 
\rangle \times \langle R_o \rangle$ space, which is shown in Figure~\ref{fig:beta_Ro}.  The NPN and PNP groups
are now much more dispersed, but remaining almost completely separated.  Interestingly, the parameter variations
for these groups in the $\langle beta \rangle \times \langle R_o \rangle$ are more comparable one another, though
the PNP is still more scattered than the NPN.  The latter type of transistors also resulted with higher $\langle R_o \rangle$,
while the $\langle \beta \rangle$ values of the two groups have strong overlap, with comparable averages.
Also, the subgroups corresponding to the 14 types of transistors are much more separate in this space than in the
Early mapping, corroborating previously similar results obtained for NPN devices~\cite{costafeed:2017}.  
Most of these subgroups also
tend to present a similar orientation, being in agreement with the similar cluster orientations
observed for NPN BJTs in~\cite{costafeed:2017}.

\subsection{The Prototypical NPN-PNP Early Space}

By combining the experimental results and analytical expressions derived in this work with pattern recognition
concepts, namely the supervised method of Linear Discriminant Analysis -- 
LDA(e.g.~\cite{johnson:2002,shapebook,costacomp:2014}), it has been 
possible to derive a  prototype NPN-PNP Early space.  As it will be shown, this prototype space shows with simplicity 
and objectively  the overall properties and parameter variation of these devices.

We start by obtaining isoline curves in the Early space $(Va, s)$ by using Equation~\ref{eq:iso}, derived from Equation~\ref{eq:rels}.

\begin{equation}
     s_{\langle \beta \rangle} = \frac{6 \langle \beta \rangle V_{C,max}}{ 2 I_{C,max}^2 ln \left( \frac{V_a-V_{C,max}}{V_a} \right) + 3 V_{C,max}^2 - 6 V_a V_{C,max} }.
     \label{eq:iso}
\end{equation}

\vspace{0.5cm}

By using this equation, it is possible to draw level-sets of fixed $\langle \beta \rangle$ in the Early space $(V_a,s)$. 
Also, we apply LDA in order to find the optimal linear separation border between the NPN and PNP groups of 
considered BJTs.  LDA is a multivariate statistics-based methodology capable of identifying, given a dataset
of objects and respective measurements and categories, the orientations along which the categories are maximally
separated according to a criterion based on the scattering of the overall and category measurements.  

\begin{figure*}[h!]
\centering{
\includegraphics[width=14cm]{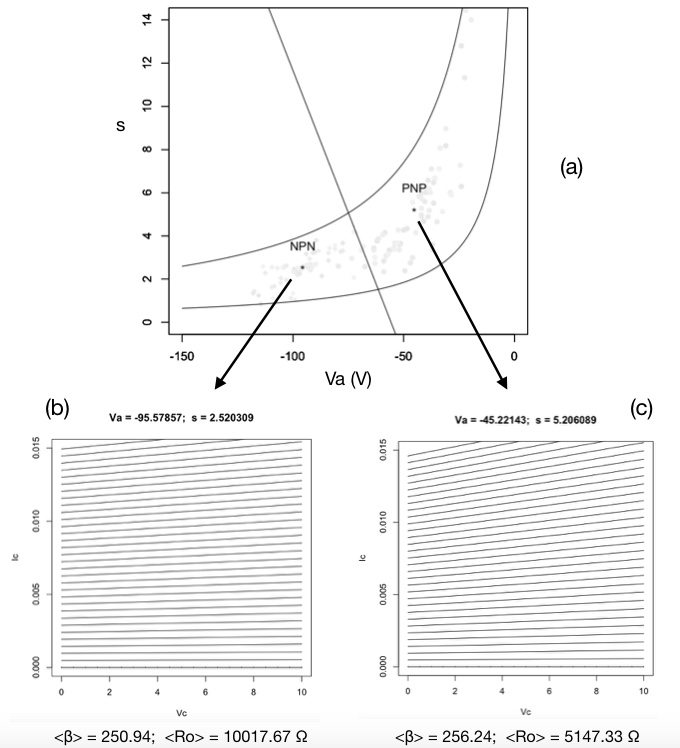}
\caption{The prototypical NPN-PNP Early Space. Also shown are the average values of $V_a$ and $s$ for the
NPN and PNP BJTs, the respective $\langle \beta \rangle$ and $R_o$, as well as the optimal separatrix
(according to linear discriminant analysis) between the two groups.  The the real-world transistors are completely
comprised between the relatively narrow curved region defined between the $\langle \beta \rangle = 100$  and 
$\langle \beta \rangle = 100$ isolines.  The real-world positions are also showed in diluted gray levels.  }
\label{fig:prot}}
\end{figure*}

The obtained results are shown in 
Figure~\ref{fig:prot}(a), which also includes a diluted representation
of the devices.  Remarkably, all the considered transistors resulted between the two isolines, which correspond to
$\langle \beta \rangle = 100$  (lower curve) and $\langle \beta \rangle = 400$ (upper curve).  The oblique line 
corresponds to the separatrix between the two groups according to the LDA optimality criterion (based on the
distance and dispersion of the features of the devices n the two groups).  The average values of $V_a$ and $s$ are
identified for each group.  The NPN group has overall variance (trace of the respective intra-class dispersion 
matrix~\cite{shapebook}) equal to  $8686.01$ while the dispersion obtained for the PNP devices is equal to
$12523.82$.  This indicates that the used PNP devices have considerably large parameter variation than the 
NPN transistors.  

The centers of mass (averages) of the
$V_a$ and $s$ of the NPN and PNP groups are also identified in this figure, and the respective theoretical
isolines characterizing the electronic operation of the NPN and PNP prototypical devices in the linear operation
region are shown as (b) and (c), respectively.  Observe the markedly more inclined isolines obtained for the
PNP transistors, implying in smaller output resistance.  Interestingly, these two prototype 
characteristic spaces are characterized by similar $\langle \beta \rangle$.  However, the prototype PNP
device has about twice as large output resistance $R_o$. 

\subsection{Theoretical THD Analysis Reveals a Surprising Feature of BJTs}

In this section, the obtained Early modeling analytical relationships are considered in a theoretical
THD analysis.  More specifically, we aim at quantifying THD levels in terms of the Early parameters
$V_a$ and $s$, as well as their mapping into the $\langle \beta \rangle \times \langle R_o \rangle$ space.

We start by deriving some general results regarding THD of signals given in terms of input and output functions.   
Let the input signal be a perfect sinusoidal $x(t) = A_x sin(2 \pi f_o t)$,
where $A_x$ is the signal amplitude, $f_o$ is its frequency, and $t$ is time.  In order to determine the THD implied
by a system that produces an output signal $y(t) = g(x(t))$, where $g()$ is a generic function, we obtain
$g(x(t))$, calculate its Fourier transform, and apply the traditional THD definition:

\begin{equation} 
  T\!H\!D(g(x(t))_{f_o} = \frac{\sqrt{S^2_{2{f_o}}  + S^2_{3{f_o}} + \ldots}} {S_{f_o}} ,  \label{eq:Ic_sin}
\end{equation}

where $S_{f_n}$ is the RMS voltage of the $n-th$ harmonic of $g(x(t))$.

Now, the first interesting property to notice is that the THD does not depend on constant terms ($b$) or proportionality
constants ($a$), i.e. 

\begin{equation}
  THD(g(x(t))) = THD(a g(x(t) + b)). \label{eq:propaffine}
\end{equation}

The  second interesting property of THD is that it does not depend on time scalings of the original
signal $x(t)$, i.e.:

\begin{equation}
  THD(g(x(t))) = THD(a g(x(r t)), \label{eq:propscaling}
\end{equation}

provided complete periods of $x(t)$ are used in both cases.

Now, it has been shown~\cite{costaearly:2017} that, for a common emitter with null emitter resistance and the load 
attached between the collector and $V_{CC}$ of an NPN transistor, we have:

\begin{eqnarray} 
  V_C(t) = \frac{V_{CC} + R_L V_a tan(s I_B)}{R_L tan(s I_B) + 1}   \label{eq:Vc} \\
  I_C(t) = \frac{(V_{CC} -V_a)tan(s I_B)}{R_L tan(s I_B) + 1} .  \label{eq:Ic}
\end{eqnarray}

A third equation can be added to express the power dissipated:

\begin{equation} 
  P_C(t) = V_C(t) I_C(t) =  \frac{V_{CC} tan(s I_B) (V_{CC}-Va)}{\left[ R_L tan(s I_B +1) \right] ^2}  \label{eq:Pc}. \\
\end{equation}

The THD implied by the mapping of $I_B$ into $I_C$ can be expressed by using
Equation~\ref{eq:Vc} and imposing $x(t) = I_B(t) = A_x sin(2 \pi f_o t)$:

\begin{equation}
  T\!H\!D_{f_o} \left( \frac{(V_{CC} -V_a)tan(s A_x sin(2 \pi f_o t))}{R_L tan(s A_x sin(2 \pi f_o t)) + 1} \right).
\end{equation}

By considering the property in Equation~\ref{eq:propaffine}, we have:

\begin{equation}
  T\!H\!D_{f_o} \left( \frac{1}{R_L  + 1/tan(s A_x sin(2 \pi f_o t))} \right).
\end{equation}

Surprisingly, it can be immediately verified that the current THD of the considered common emitter NPN 
configuration operating in linear region (Class A), \emph{does not depend on $V_a$, but only on $s$}.  
Property in Equation~\ref{eq:propscaling}
extends this remarkable result to any input frequency $f_o$, considering absence of reactive effects.

\begin{figure*}[h!]
\centering{
\includegraphics[width=15cm]{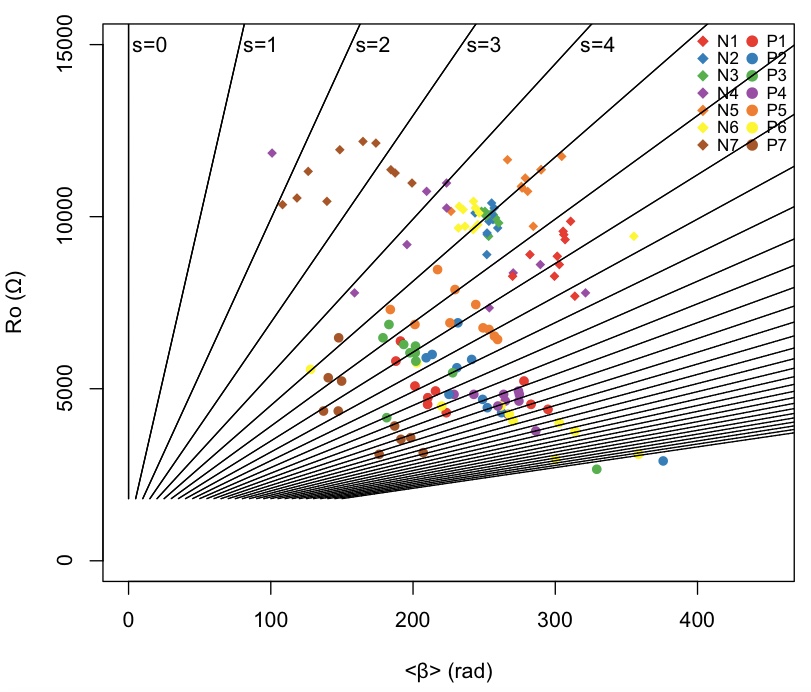}
\caption{The lines defined in the $\langle \beta \rangle \times \langle R_o \rangle$ space by the lines
in the Early space such that $s=ct$ and $-150V \leq V_a <0$.  The lines are shown respectively to
$s = 0, 1, \ldots, 30$. The smallest THDs are obtained for $s$ small, so that the NPN components of the
$\pi_7$ and $pi_4$ complementary pair allow the best linearity, as far as THD is concerned, among the considered 
devices for the adopted
configuration, i.e. $V_{CC}=12V$, $I_{C,min} = 1mA$, $I_{C,max} = 15mA$ and $R_L = 700 \Omega$.}
\label{fig:anl_THD}}
\end{figure*}

This result, combined with Equation~\ref{eq:av_beta} and the application of numerical Fourier transform
over finely sampled signal versions (it does not seem possible to derive aan nalytical expression for the
THD in this case), allow us to draw the lines in the $\langle \beta \rangle \times
\langle R_o \rangle$ where the THD will be constant, as defined by fixed values of $s = 0, 1, \ldots, 30$.
Figure~\ref{fig:anl_THD} depicts the so obtained results.  Given a THD level, determined by the respective
Early parameter $s$, the transistors allowing that level will have  $\langle \beta \rangle$ directly proportional
to $\langle R_o \rangle$ with a constant of proportionality that decreases in a non-linear fashion with $s$. 
Among the considered BJTs, the most linear operation --- as far as THD in the adopted configuration is
concerned, --- is allowed by $N_7$ and $N_4$.  The PNP transistors $P_6$ tend to be particularly
non-linear in the considered circumstances.  Most of the adopted devices resulted comprised between the 
THD lines ranging from $s=4$
to $s=17$, which is a relatively large distortion range.   In other words, great distortion variation can be
potentially implied by not considering the Early parameter $s$ of the chosen transistors.

\subsection{Experimental THD Analysis with Resistive Loads}

The complementary pairs having the smallest and largest parameter differences were considered
for experiments taking into account the three push pull amplifier configurations discussed in Section~\ref{sec:pushpull}.  
These two pairs of transistors are $(N_4(9),P_5(9))$ (smallest parameter difference) and $(N_7(8),P_3(7))$, respectively.
A purely resistive load $R_L = 670 \Omega$ was adopted throughout, and $V_{CC} = 12V$.  The total harmonic
distortion, THD (e.g.~\cite{cordell:2011}), was estimated by using a high quality THD analyzer.   The input signals
consisted of purely sinusoidal waves with peak-to-peak amplitudes $V_i = 1, 2, 3, 4, 5V$.  Table~\ref{tab:THDs}
presents the THD values (in \%) obtained for each configuration of transistor pair and push pull circuit.  As could be expected,
considerably large THD values were obtained as a consequence of the simplicity of the considered circuits.

In the case of the simplest push pull configuration, implying large cross-over distortion,  the THD values tend to decrease with
the input voltage amplitude, because the cross-over effect becomes relatively smaller with respect to the total
output voltage.  As expected, the THD tends to increase with $V_i$ in the diode  push pull configuration.  A
variable trend was observed for the negative feedback circuit.  

Remarkably, and with only two exceptions, the THD values obtained in the three types of push pull circuits were
smaller in the BJT pair with more similar parameters, namely $(N_4(9),P_5(9))$.  The exceptions were observed
for the negative feedback configuration for $V_i = 4$ and $5V$.   These results tend to confirm two important trends,
namely that better results are obtained with complementary pairs having similar parameters, and that negative
feedback may not be enough to completely eliminate the effects of parameter variability~\cite{costafeed:2017}.  

\begin{table}
  \centering
     \begin{tabular}{|| c || r | r | r | r | r || }
        \hline
        Pair (push pull config.)& 1V & 2V & 3V & 4V & 5V  \\
        \hline \hline
      $(N_4(9),P_5(9))$  (simple)     & 32.03 & 15.59 & 10.38 & 7.01 & 5.62 \\ 
      $(N_7(8),P_3(7))$  (simple)      & 34.05 & 17.15 & 11.15 & 7.83 & 6.25   \\ 
      $(N_4(9),P_5(9))$  (diodes)      &  0.27 &  0.56   &  0.69 &  0.72 &  0.73  \\ 
      $(N_7(8),P_3(7))$  (diodes)      &  0.62 &  1.03   &  1.15 &  1.05 & 1.03  \\ 
      $(N_4(9),P_5(9))$ (neg. feed.) &  0.29 &  0.82  &   0.59 &  0.64 & 0.65  \\ 
      $(N_7(8),P_5(7))$ (neg. feed.) &  0.36 &  0.91 &    0.68 &  0.61 & 0.62  \\ 
      \hline
      \end{tabular}
    \caption{The experimental THD values (in percentage) obtained for the transistor pair $(N_4(9),P_5(9))$, having the closest
    parameters among all possible pairwise combinations, and the pair $(N_7(8),P_3(7))$, characterize by 
    exhibiting the most discrepant parameters. Each of these pairs was used in each of the three considered
    push pull configurations with $R_L=670\Omega$ and $V_{CC} = 12V$.  Smaller THD values are obtained in 
    almost all respective configurations.}
    \label{tab:THDs}
\end{table}

\begin{figure}[h!]
\centering{
\includegraphics[width=8cm]{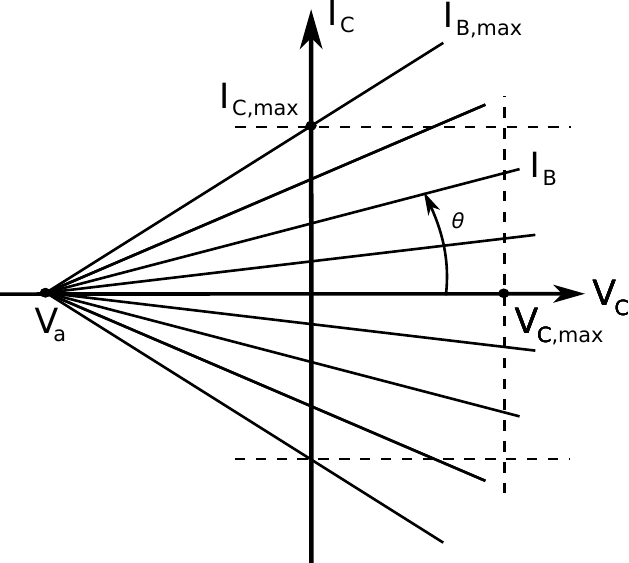}
\caption{The Early model geometric configuration as extended for joint operation of a complementary pair of BJTs.}
\label{fig:Earlypp}}
\end{figure}

It is interesting to observe that push pull configurations can be conceptualized as a single device
in the Early modeling, which is achieved by reflecting the PNP portion and merging it into the
NPN Early diagram in Figure~\ref{fig:Early}.  This combined representation is illustrated in Figure~\ref{fig:Earlypp}.
The use of complementary pairs with unmatched parameters will imply the two quadrants respective to
the NPN and PNP operation to be asymmetric, resulting in larger distortions, as observed in the reported
experiments.

\section{Concluding Remarks}

Reproducibility, stability, and linearity are most desired properties in BJT amplifying circuits. 
The former property is limited by the inherent large variability of the characteristic parameters 
of junction transistors, and the latter is a consequence of several effects, including the dependence
of the transistor parameters with the operating voltages and currents, intrinsic non-linearities implied
by quantum mechanics, as well as circuit design issues (e.g. the cross-over distortion in
push pull configurations).  Since the introduction of BJTs in the mid 40's, much effort has been invested
in better understanding and controlling the intrinsic properties, often expressed as parameters, of
this important class of transistors, responsible for the onset of modern Electronics.   Many models
of BJTs and circuit configurations have been proposed, several of them involving the current
gain $\beta$ or $h_{fe}$ and the collector output resistance $R_o$.  Despite corresponding to
quantities of more intuitive applicability and significance, these parameters have
an intrinsic limitation in the sense that they constantly depend on the $V_c$ and $I_C$ even under
normal circuit operation.  So, oftentimes BJTs are characterized in terms of \emph{averages} of these
parameters, or values taken at a specific operation point. 

One of the resources that has been applied in order to reduce power consumption in BJT amplifiers
involve the application of a complementary pair operating in push pull configuration classes B or AB.
This operation can imply in cross-over distortion, which has to be controlled by some effective means
such as diode-induced base biasing and negative feedback.  The push pull configuration is particularly
appealing from the theoretical point of view because of its symmetry and complementarity, which
characterize the unified operation of this configuration.  This configuration is particularly
important because it is often employed as output stage of many off-the-shelf operational amplifiers.  
One particularly interesting issue with push pull operation is the inherent difference between NPN
and PNP devices, which is mostly a consequence of different mobilities in the N and P semiconductors,
and geometrical parameters~\cite{parker:2004,boylestad:2008}.  Therefore, it would be interesting to achieve
a better understanding, if possible in simple and intuitive fashion, of the characteristics and variability intrinsic 
to each of the two main transistors types.  Actually, NPN BJTs tend to be more often used in practice than the 
PNP counterparts as a consequence of differentiating properties between these two categories. 

Advances in instrumentation, pattern recognition, image analysis, and transistor modeling have jointly paved 
the way to more comprehensive integrated studies of semiconductor devices and circuits.  In particular,
a recently introduced Early modeling for BJTs~\cite{costaearly:2017} provided a simple and intuitive framework for
investigating devices and circuit configurations.  Of special interest is the fact that the two main involved
parameters, namely the Early voltage $V_a$ and the proportionality parameter $s$ largely do not depend
on $V_c$ and $I_C$ during normal circuit operation, being almost constant and intrinsic to each individual
BJT.  In addition, the Early model provides a simple and elegant geometrical understanding of BJT
operation and parameters.  In the present work, we used an improved version of the Early modeling
approach reported in\~cite{costaearly:2017} in order to investigate the properties of complementary pairs of BJTs.  
In particular, the estimation of $V_a$ is performed by a voting scheme motivated by the image analysis
technique known as the Hough transform, capable of reducing the interference between the $I_B$ isolines.
This modification corresponds to the first main contribution of the present work.  In addition, we
derive relationships between the Early parameters $V_a$ and $s$ with the more traditionally applied
current gain $\beta$ and output resistance $R_o$.  Because these two latter parameters depend on
$V_C$ and $I_C$, their average value in the operation region had to be taken into account in the
derived relationships with the Early parameters.  These relationships, which can be adapted to more
general operation regions than the first quadrant $Q$, constitute another contribution of the present
article, which allowed the Early approach and results to be related and interpreted in terms of the more
traditional and intuitive parameters $\beta$ and $R_o$.  The whole procedure for estimating the Early
parameters, including signal capture, pre-processing, adherence residuals, as well as the numeric 
estimation procedure were described, discussed and illustrated.  The acquisition system, build
exclusively for the reported experiments, was also briefly described.

Th reported theoretic-experimental framework was then applied for the characterization of 7 types of 
complementary BJT pairs, each represented by 10 respective samples, totaling 140 real-world small signal 
devices.  Several interesting results were identified and discussed.  When mapped onto the Early
space, the real-world devices resulted comprised within a relatively narrow band with increasing values
of $s$ for larger values of $V_a$.  The NPN and PNP groups of devices resulted markedly separated
in the Early space, with the former exhibiting improved linearity at the cost of larger output resistance.
The PNP group was also characterized by larger parameter variation, especially in which concerns
the proportionality parameter $s$.   The region defined by bounding isolines of $\langle \theta \rangle$
for constant values of 100 and 400, obtained by using the theoretically derived relationship between
the Early parameters and the more traditionally used $\beta$, was found to neatly contain all the
experimental devices mapped in the Early space.  These results regarding the two main transistor types 
constitute another of the main contributions of the present work.  

The variability and relationships between the other parameters, namely $R_o$, $cs$ and $I_a$, were also 
considered and discussed.  By incorporating linear discriminante analysis,
it was possible to identify an optimal separatrix between the NPN and PNP groups.   A prototype
NPN-PNP Early space was then derived by combining this separatrix with the
100 and 400 $\beta$ isolines, and the average Early parameters.  This space summarizes in an
effective and yet comprehensive way the main properties and respective variability of the two main types of 
BJTs.   The theoretical isolines corresponding to the averages obtained for the NPN and PNP groups were
also derived and illustrated, corresponding to prototypical isoline respective configurations in the $(V_C,I_C)$ space.
Remarkably, the obtained prototype NPN and PNP devices resulted with similar $\langle \beta \rangle$ values,
but considerably different $V_a$ and $s$, as well as $\langle R_o \rangle$.  This suggests that NPN and PNP
tend to be, in the average, matched regarding the current gain $\beta$.  The divergence of characteristics
would therefore be accounted by differences between the other parameters, such as $s$.  The obtained 
relationships between the Early parameters and $\langle \beta \rangle$ and $\langle R_o \rangle$ were then 
used to map the considered real-world BJTs into the space defined by the latter pair of parameters.   The 
results are particularly interesting, with the two groups resulting again almost completely apart, and with  
PNP still exhibiting larger parameter variability, though to a smaller extent, than NPN
counterparts.  In addition, as in~\cite{costafeed:2017}, well-defined groups were obtained for each of the
14 experimentally characterized transistor types, which substantiates the fact that the variability within each
transistor type tends to be smaller than the variation between the different types.  These results also indicated
that the similar cluster orientations obtained in~\cite{costafeed:2017} are defined by the narrow curved band
occupied by the respective devices in the Early space.

The obtained relationship between the Early parameters and the more commonly used $\beta$ and $R_o$ 
parameters also allowed the THD of BJTs to be analytically investigated.  By using THD properties and
the obtained relationships, considering a simplified common emitter configuration~\cite{costaearly:2017},
a remarkable result, which constitutes another of the main contributions of the present work, was achieve 
in which the THD, for a fixed circuit configuration, \emph{does not depend on 
$V_a$, but only on $s$}.   By mapping the lines defined in the Early space for $s=ct.$, it was possible to
identify that oblique lines with positive slopes are defined in the more traditional $\langle \beta \rangle \times
\langle R_o \rangle$ space.  Thus, for a fixed THD level indexed by $s$, if a transistor with that level
has higher $\beta$, it will have proportionally higher $R_o$.  The considered BJTs were observed to have
their THD levels to vary significantly among the samples of the same transistor type, with interesting
practical implications.

To conclude the reported investigation, two hybrid (in the sense of involving transistors from different
types) complementary pairs, one with the closest parameter configuration and the other with the most
diverging parameters, were taken back to the bench and used in three push pull configurations
amplifying sinusoidal signals with progressively increasing voltage amplitudes onto a resistive load.  
In the greatest majority of cases, the use of the 
better matched pair led to improved THD values, even in the presence of negative feedback.  These results 
corroborate, at least for the considered devices and circuits, the validity of using complementary transistors with
similar parameter values whenever possible.   Interestingly, it turns out to be a particularly challenging
task because, as indicated in the reported experimental results, NPN transistors tend to occupy rather
distinction positions in the Early space, implying in markedly divergent electronic properties.  

The efficacy of the reported methodology, as well as the several results obtained regarding NPN
and PNP characteristics and parameters variation pave the way to a number of future related works,
which are discussed as follows.

First, it would be interesting to extend theoretically and experimentally the reported approaches
to other families of transistors, such as FET and MOS, as well as to high frequency and high power
BJTs.  Also promising would be to experimentally map older device technologies, in particular those based on 
germanium, in order to compare their parametric features with the more 
modern silicon counterparts.   Preliminary experiments performed by the author, considering an old new stock
(NOS) lot of a type of NPN germanium junction transistors, suggested that these devices are characterized
by very small $s$ values and small $V_a$ values, the latter being comparable to those of silicon PNP devices.   
Another intriguing question concerns if the Early approach can be eventually applied to other types of amplifiers, 
such as optical, mechanical, acoustical, etc.

Regarding complementary pairs, it would be of interest to consider, in a similar
fashion as in~\cite{costaarray:2017}, off-the-shelf transistor
arrays containing matched NPN and PNP transistors.  Further investigations are also required in order
to possibly extend the obtained results to other models of transistors operating in different circuit
configurations, not to mention transistor arrangements such as Darlington tandem and in parallel.  
Also, the proposed methodology can be applied to infer the effects of negative feedback in the
resulting Early parameters, such as in common emitter and push pull configurations.  Of special
interest would be the consideration of reactive loads typically found in practice.  It is believe that
the results reported presently can be directly extended to this type of loads.

It would also be interesting to consider the here identified parametric relationships
in terms of conformal mappings ~\cite{churchill:1984}.   
It is also believed that, with simple adaptations, the proposed framework 
can be extended to the characterization and analysis of sensors and transducers.  
In this respect, it is believed that interesting experiments can be performed with off-the-shelf
optocouplers.  Needless to say, all the concepts, methods and results
presented in this work can be natural and effectively extended to transistors in integrated circuits at
their progressive integration levels.  Also, It would be worth investigating in which sense
the results reported in this work could be applied to digital electronics, especially concerning the
switching time.

At the level of the physics of semiconductors devices, it would
be promising to use the reported methods in order to identify possible relationships between the 
Early parameters and other properties characteristics of PNP and NPN materials.  In particular,
it would be to check the effects of temperature over the Early parameters.  

Several of the above mentioned venues for further research are being currently investigated and 
results should be reported opportunely.  It should be also observed that the Early approach to 
modeling and characterizing transistors is particularly simple and has good potential for being
used for didactic purposes regarding device and circuit theory and practice.

All in all, the Early effect provides a simple and efficient geometrical framework for approaching
analog electronic devices and circuits, ranging from discrete devices to VLSI.  It remains an
additional interesting question, worth of further study, wether the convergence of isolines observed
for BJTs (and also other families such as MOSFETs) constitutes a type of universal feature exhibited
by most amplifying and modulating devices of several natures and technologies..

\vspace{0.7cm}
\textbf{Acknowledgments.}

Luciano da F. Costa
thanks CNPq (grant no.~307333/2013-2) for sponsorship. This work has been supported also
by FAPESP grants 11/50761-2 and 2015/22308-2.
\vspace{1cm}

%\nocite{*}

\bibliography{mybib}
\bibliographystyle{plain}
\end{document}